\def\preprint{1}		

\ifdefined\preprint
  \documentclass[preprint,review,12pt]{elsarticle}
\fi
\ifdefined\wordcount
  \documentclass[final,3p,times,twocolumn]{elsarticle}
\fi
\ifdefined\final
  \documentclass[final,3p,times,twocolumn]{elsarticle}
\fi

\usepackage{lineno,hyperref}
\usepackage{placeins}
\usepackage{graphicx}
\usepackage{adjustbox}
\usepackage{subfigure}
\usepackage{color}
\usepackage{amsmath}
\usepackage{amssymb}
\usepackage{physics}
\usepackage{wrapfig}
\usepackage{chemformula}
\usepackage{placeins}
\usepackage{siunitx}
\journal{Fuel}

\biboptions{sort&compress}

\begin{document}

\begin{frontmatter}

\title{Flashback propensity due to hydrogen blending in natural gas: sensitivity to operating and geometrical parameters}


\author[1]{Filippo Fruzza}
\author[1]{Rachele Lamioni\corref{mycorrespondingauthor}}
\cortext[mycorrespondingauthor]{Corresponding author}
\ead{rachele.lamioni@unipi.it}
\author[1]{Alessandro Mariotti}
\author[1]{Maria Vittoria Salvetti}
\author[1]{Chiara Galletti}
\address[1]{Dipartimento di Ingegneria Civile e Industriale, Università di Pisa, 56122 Pisa, Italy}

\begin{abstract}
Hydrogen has emerged as a promising option for promoting decarbonization in various sectors by serving as a replacement for natural gas while retaining the combustion-based conversion system. However, its higher reactivity compared to natural gas introduces a significant risk of flashback. This study investigates the impact of operating and geometry parameters on flashback phenomena in multi-slit burners fed with hydrogen-methane-air mixtures. For this purpose, transient numerical simulations, which take into account conjugate heat transfer between the fluid and the solid walls, are coupled with stochastic sensitivity analysis based on Generalized Polynomial Chaos. This allows deriving comprehensive maps of flashback velocities and burner temperatures within the parameter space of hydrogen content, equivalence ratio, and slit width, using a limited number of numerical simulations. Moreover, we assess the influence of different parameters and their interactions on flashback propensity. The ranges we investigate encompass highly \ch{H2}-enriched lean mixtures, ranging from 80\% to 100\% \ch{H2} by volume, with equivalence ratios ranging from 0.5 to 1.0. We also consider slit widths that are typically encountered in burners for end-user devices, ranging from ${\SI{0.5}{mm}}$ to ${\SI{1.2}{mm}}$. The study highlights the dominant role of preferential diffusion in affecting flashback physics and propensity as parameters vary, including significant enrichment close to the burner plate due to the Soret effect. These findings hold promise for driving the design and optimization of perforated burners, enabling their safe and efficient operation in practical end-user applications.
\end{abstract}

\end{frontmatter}
\section{Introduction}

The recent EU Fit for 55 program emphasizes the urgency of reducing carbon dioxide (\ch{CO2}) emissions by 55\% by 2030 compared to 1990 levels, with the ultimate goal of achieving a carbon-neutral Europe by 2050~\cite{FitTo55}. Meeting this target necessitates a significant increase in the use of renewable energy sources to minimize reliance on fossil fuels, as outlined in the REPowEREU action~\cite{rePower}. The residential sector accounts for approximately 30\% of final energy consumption, and by 2050, at least two-thirds of the energy demand in Europe must be fulfilled through electricity to achieve the aforementioned objectives. However, the current heating systems heavily depend on gas boilers, making the transition to electrification a costly and challenging endeavor~\cite{QUARTON2020115172}. To further accelerate decarbonization, the International Energy Agency has recommended that only zero carbon-ready gas boilers be available for sale after 2023~\cite{mouli2021comparative,YANG2023}.

In this scenario, an interesting option is provided by hydrogen produced via water electrolysis, fed by excess wind and solar power~\cite{YOUNAS2022123317}. Indeed, hydrogen can be utilized as a green energy carrier for gas boilers, employing similar technologies and minimizing the expenses associated with heating system upgrades~\cite{MCKENNA2018386_1,michalski2017hydrogen}, as sometimes electrification is hard to achieve because of economic reasons and/or other constraints (for instance in historical buildings). Additionally, hydrogen will soon be introduced into existing gas networks at concentrations of up to 20\%, enabling the storage and distribution of renewable energy through the current infrastructure~\cite{SMITH202223071,ZHAO201912239,DEVRIES2020114116,WU20228071} to end-user devices. Furthermore, there are expectations for the development of pipelines capable of accommodating 100\% hydrogen, especially with the establishment of regionally integrated hydrogen ecosystems known as ``Hydrogen Valleys", albeit this transition will require substantial efforts and occur gradually~\cite{KOVAC202110016}. It is projected that by 2050, hydrogen, either as a hydrogen-gas mixture from the gas network or in its pure form within hydrogen valleys, could satisfy up to 18\% of the energy demand for residential heating~\cite{h2europe}.

The key issue is that the replacement of natural gas with hydrogen should be accomplished safely by preserving the required performances in terms of efficiency and pollutant emissions of domestic condensing boilers. These devices typically employ premixed cylindrical or flat perforated burners, as described by Najarnikoo et al.~\cite{Najarnikoo2019}, Lamioni et al.~\cite{LAMIONI2023IJHE,LAMIONI2022CTM}, Schiro et al.~\cite{schiro2019experimental} and Jithin et al.~\cite{EdacheriVeetil2018}. The burner design is well-consolidated, thanks to the extensive manufacturers' experience, for natural gas~\cite{schiro2019experimental}. However, the physical properties of hydrogen differ strongly from those of natural gas, thus we have to understand the impact of using \ch{H2}-enriched mixtures, up to 100\% \ch{H2}, on the combustion process and eventually devise modifications of the burner design. Using hydrogen can lead to excessive mixture temperatures, which may compromise the device's performance. Furthermore, the laminar flame speed of hydrogen can exceed that of natural gas by more than six times, posing challenges when utilizing hydrogen in practical applications~\cite{KONNOV20101392} with difficulties in achieving flame stabilization. Indeed, flame stability relies on the local balance between flow velocity and flame front propagation velocity along the flame front: if the flow velocity falls below the flame front propagation velocity, the flame front moves upstream in search of a new stable configuration. However, when a stable configuration cannot be attained, the flame propagates towards the burner and upstream components, resulting in an undesired phenomenon known as flashback~\cite{KURDYUMOV20001883,SU201421307,JIMENEZ201512541}. Several factors significantly contribute to flame stabilization, including flame-wall conjugate heat transfer, flow-flame interaction, curvature, stretch rate, Soret diffusion, and preferential diffusion effects~\cite{ZHANG202010920,KEDIA20121055,VANCElewis,VANCEsoret,WANG2020117051,JIN2022122292}. Notably, interactions between the flame and the walls play a crucial role in the flashback dynamics. Kurdyumov et al. emphasized the influence of wall heat losses by comparing conditions with and without wall insulation~\cite{KURDYUMOV20071275}. Kiymaz et al. investigated the effects of wall temperature on flashback susceptibility in Bunsen-type burners, emphasizing the importance of heat transfer between the flame and the solid burner~\cite{KIYMAZ202225022}. Xia et al. conducted a numerical investigation into the influence of thermal boundary conditions on boundary layer flashback in a bluff-body swirl burner using a \ch{H2}-\ch{CH4}-air mixture~\cite{XIA20234541}. Additionally, the significant effect of flow-flame interactions on flashback dynamics in hydrogen-enriched swirled flames was recently emphasized by Ebi and Clemens~\cite{EBI201639} and Ranjan and Clemens~\cite{RANJAN20216289}. These studies collectively shed light on the complex interplay between various factors and their impact on flame stabilization and flashback phenomena.

Indeed, flashback is one of the main concerns hampering the implementation of hydrogen in domestic condensing boilers. As these devices typically operate with a premixed configuration, the potential occurrence of flashback may pose hazardous situations~\cite{BOULAHLIB202137628}. Thus, it is crucial to carefully assess the implications of incorporating \ch{H2} into domestic devices and develop designs that effectively prevent flashback phenomena. Recently, flashback limits of \ch{H2}-enriched mixtures in domestic boilers have been investigated experimentally evaluating the impact of wall temperature~\cite{deVries2017impact,ANIELLO2022}, and exploring the potential of autoignition as a mechanism for flashback initiation~\cite{PERS2022}. On the other hand, some correlations based on numerical models have been proposed to estimate the flashback velocities of hydrogen in practical configurations. Vance et al.~\cite{VANCEcorrelation} investigated the flashback limits of premixed hydrogen flames in burners with multiple slits, considering both fluid and solid regions and incorporating conjugate heat transfer in their analysis. In a recent paper from our group, we developed a transient model capable of estimating flashback limits and burner temperatures of \ch{H2}-\ch{CH4}-air mixtures in multi-slit configurations, identifying different flashback regimes depending on the hydrogen content~\cite{FRUZZA2023}.  
However, a comprehensive understanding of the relative importance and interplay between the various parameters, including burner design and operating conditions, is still lacking. This comprehension is necessary for the identification of practical solutions to implement the addition of hydrogen to the gas mixture efficiently. In this perspective, this study investigates numerically the effects of various parameters on the flashback propensity of \ch{H2}-\ch{CH4}-air mixtures. We employ generalized Polynomial Chaos (gPC) expansion~\cite{Ale1} to construct a comprehensive map of critical flashback velocities and burner temperatures. This approach allows us to explore the parameter space encompassing hydrogen content, equivalence ratio, and geometry parameters using a limited number of Computational Fluid Dynamics (CFD) simulations. In addition, the gPC-based stochastic sensitivity analysis helps to quantitatively understand the influence of the different parameters and their interactions on the flashback propensity, assessing the relative importance of each parameter and gaining insights into their combined effects. The simulations are conducted using a 2D transient numerical model that replicates an array of slits found in real perforated burners commonly used in condensing boilers. The interpretation of the results is guided by the underlying physics governing the flashback phenomenon. We discover that preferential diffusion effects, deriving from different sources, significantly influence the flashback limits by altering the flashback physics. These effects play a crucial role and must be carefully considered to achieve a comprehensive understanding of flashback dynamics in \ch{H2}-\ch{CH4} mixtures.

\section{Numerical model}

\begin{figure}[!htbp]
\centering
\includegraphics[width=0.8\textwidth]{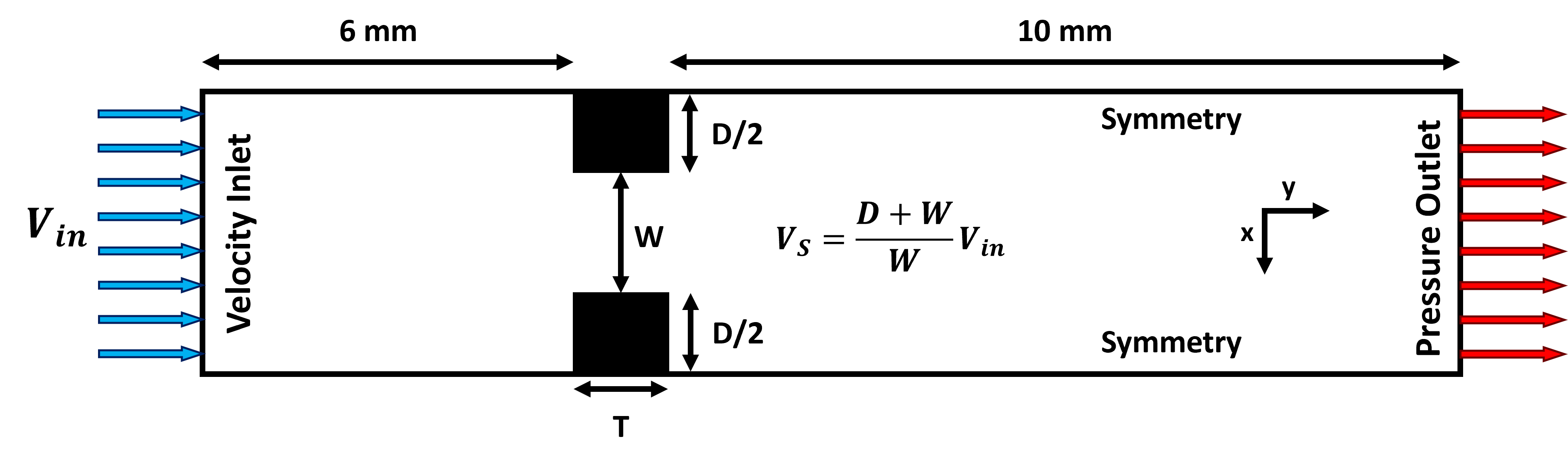}
\caption{Computational domain.}
\label{fig:domain}
\end{figure}

We consider a multi-slit configuration representing a section of an actual burner commonly employed in residential condensing boilers. The estimation of flashback velocity within this particular configuration is accomplished through the numerical model and procedure suggested in Fruzza et al.~\cite{FRUZZA2023}. Taking advantage of symmetry conditions, the computational domain comprises a single infinitely-long 2D slit as shown in Fig.~\ref{fig:domain}, where the solid zone corresponding to the burner is represented in black. The slit width is denoted by $W$, while the distance between two adjacent slits is denoted by $D$. The slit width is varied in the range ${W \in [\SI{0.5}{\mm},\SI{1.2}{\mm}]}$. For any variation of $W$, the distance between the slits $D$ is varied to keep fixed the porosity of the burner at ${W/D = 0.5}$. The burner thickness is ${T=0.6}$ mm for every case. The domain extends enough both downstream (${\SI{10}{\mm}}$) and upstream (${\SI{6}{\mm}}$) of the solid to be able to enclose the flame even in flashback conditions and hence avoid the influence of the boundary conditions on the solution.  We define the \emph{cold-flow} bulk velocity at the slit entry as 
\begin{equation}\label{eq:defv}
    V_S\equiv\frac{D+W}{W}V_{in},
\end{equation}
where $V_{in}$ is the uniform inlet velocity. This is the velocity we would have at the slit entry for a cold flow, i.e., neglecting the density variations of the mixture due to the high burner temperatures. By selecting a specific porosity value of ${W/D = 0.5}$, we maintain a consistent relationship between the inlet velocity and the slit velocity, specifically $V_S = \frac{3}{2} V_{in}$. It is important to highlight that by keeping the porosity constant, we ensure that, for a given mixture, the same $V_S$ corresponds to the identical thermal power delivered to an ideal burner completely covered by the simulated slits.

\subsection{Physical model}

The problem may be described by conservation equations for mass, momentum, energy, and transport/reaction equations for the chemical species in the gas phase:
\begin{gather}
    \frac{\partial\rho}{\partial t}+\mathbf{\nabla}\cdot\left(\rho \mathbf{v}\right)=0 \\
    \frac{\partial}{\partial t}\left(\rho \mathbf{v}\right)+\mathbf{\nabla}\cdot\left(\rho \mathbf{v}\mathbf{v}\right)=-\mathbf{\nabla}p+\mathbf{\nabla}\cdot\left(\mathbf{\Bar{\tau}}\right) \\
    \begin{split}
    & \frac{\partial}{\partial t}\left(\rho E\right)+\mathbf{\nabla}\cdot\left(\mathbf{v}\left(\rho E+p\right)\right)= \\
    & =\mathbf{\nabla}\cdot\left(k \mathbf{\nabla}T+\sum_{j=1}^N h_j \left(\sum_{k=1}^{N-1} \rho D_{m,jk}\mathbf{\nabla}Y_k+D_{T,j}\frac{\mathbf{\nabla }T}{T}\right)\right)-\sum_{j=1}^N h_j\omega_j+S_{rad}
    \end{split} \\
    \frac{\partial}{\partial t}\left(\rho Y_i\right)+\mathbf{\nabla}\cdot\left(\rho\mathbf{v}Y_i\right)=\mathbf{\nabla}\cdot\left(\sum_{j=1}^{N-1} \rho D_{m,ij}\mathbf{\nabla}Y_j+D_{T,i}\frac{\mathbf{\nabla }T}{T}\right)+\omega_i,
\end{gather}
\noindent where $\rho$ is the density, $\mathbf{v}$ is the velocity vector, $p$ is the pressure and $\mathbf{\Bar{\tau}}$ is the stress tensor. The ideal gas equation of state is used for the density. $h_i$, $Y_i$ and $\omega_i$ are the enthalpy, the mass fraction, and the net rate of production of the $j$th species respectively, and $E=\sum_{i=1}^N h_i Y_i-p/\rho+\abs{\vec{v}}^2/2$. $k$ is the mass-weighted thermal conductivity of the mixture, $D_{m,ij}$ are the generalized Fick’s law diffusion coefficients of the species $i$ in species $j$, and $D_{T,i}$ are the thermal diffusion coefficients of the $i$th species. Finally, $S_{rad}$ is the energy source associated with radiation. Inside the solid domain, we solve the energy equation:
\begin{equation}
   \frac{\partial}{\partial t}\left(\rho_s h_s\right)=\nabla\cdot\left(k_s\nabla T\right)
\end{equation}
\noindent where $\rho_s$, $k_s$, $c_{p,s}$, and $h_s=\int_{T_{0}}^T c_{p,s} dT$ are the density, the thermal conductivity, the specific heat, and the sensible enthalpy of the solid material. The equations are solved on a fixed structured grid, with characteristic cell size in the reaction front region of ${\SI{25}{\um}}$. The grid resolution was selected to guarantee a minimum of 13 cells within the flame thickness, even in the worst-case scenario with the thinnest flame. To determine this resolution, several 1D freely-propagating flames were simulated using Cantera~\cite{cantera}. We selected one of these simulations as a reference, specifically a flame with 100\% \ch{H2} at $\phi=0.8$ under the same thermodynamic conditions used in this study. By means of a grid convergence study, the solutions were proven to be well-resolved and grid-independent.  We use detailed chemistry, employing the Kee-58 mechanism with 17 chemical species and 58 reversible reactions~\cite{kee}. Full multi-component diffusion is modeled through the definition of generalized Fick's law coefficients derived by the Maxwell-Stefan equations~\cite{fluent2022,taylorkrishna1993,Merk1959}. Thermal, or Soret, diffusion is modeled using the following empirically-based composition-dependent expression provided by Kuo~\cite{kuo2005principles}:
\begin{equation}
    D_{T,i}=-2.59\times 10^{-7} T^{0.659}\left[\frac{M_{i}^{0.511}X_i}{\sum_{j=1}^N M_{j}^{0.511}X_j}-Y_i\right] \cdot \left[ \frac{\sum_{j=1}^N M_{j}^{0.511}X_j}{\sum_{j=1}^N M_{j}^{0.489}X_j} \right],
\end{equation}
\noindent where $M_i$, $X_i$, and $Y_i$ are the molar mass, molar fraction, and mass fraction, respectively, of the species $i$. Radiation is modeled by means of the gray Discrete Ordinates (DO) method~\cite{MODEST2013541}, assuming the emissivity of the fluid-solid interface to be 0.85. The burner is modeled as a solid with the properties of the stainless steel typically used for this kind of burner, with density ${\rho_s=\SI{7719}{\kg\per\cubic\metre}}$, specific heat ${c_{p,s}=\SI{461.3}{\J\per\kg\per\K}}$, and thermal conductivity ${k_s=\SI{22.54}{\W\per\metre\per\K}}$. The conjugate heat transfer (CHT) between the fluid and the solid zones is modeled to take account of the interaction between the flame and the burner plate. The CHT is modeled by employing Fourier's Law to calculate the heat flux across the fluid-solid interface cells~\cite{fluent2022}. It should be noted that no turbulence model is required, as all simulations exhibit fully laminar flow, with a maximum jet Reynolds number of 500.

\subsection{Boundary conditions}

Uniform velocity and uniform temperature of ${T_u=\SI{300}{K}}$ are set at the inlet, while a pressure outlet with ${p=\SI{1}{atm}}$ is imposed at the exit of the domain. The external edges of the domain are modeled as symmetry boundaries to take account of the interaction with the flames from the nearby slits: more specifically, we impose zero normal velocity and zero normal gradients of all variables at the symmetry plane. At the fluid-solid interfaces, a no-slip boundary condition is applied for velocity, while no thermal boundary conditions are required as heat fluxes are directly computed as described above.

\subsection{Estimation of flashback velocity}\label{sub:methodology}

To assess flashback, transient simulations are performed using the pressure-based PISO algorithm available in \textsc{ANSYS}-\textsc{Fluent 22.1}~\cite{fluent2022}. A second-order upwind scheme is utilized for both time and space discretization. Within the fluid zone, a time step of $\SI{1}{\mu s}$ is utilized. Given the significant disparity in characteristic time scales between the reactive flow in the fluid domain and the heat conduction within the solid domain, where burner temperature stabilization typically takes seconds, a time step of $\SI{1}{ms}$ is employed in the solid zone to ensure computational feasibility of the simulations. The reliability of this approach has been proven in the recent study by Fruzza et al.~\cite{FRUZZA2023}, which provides further details on the methodology.

\begin{figure}[!ht]
\centering
\includegraphics[width=0.65\textwidth]{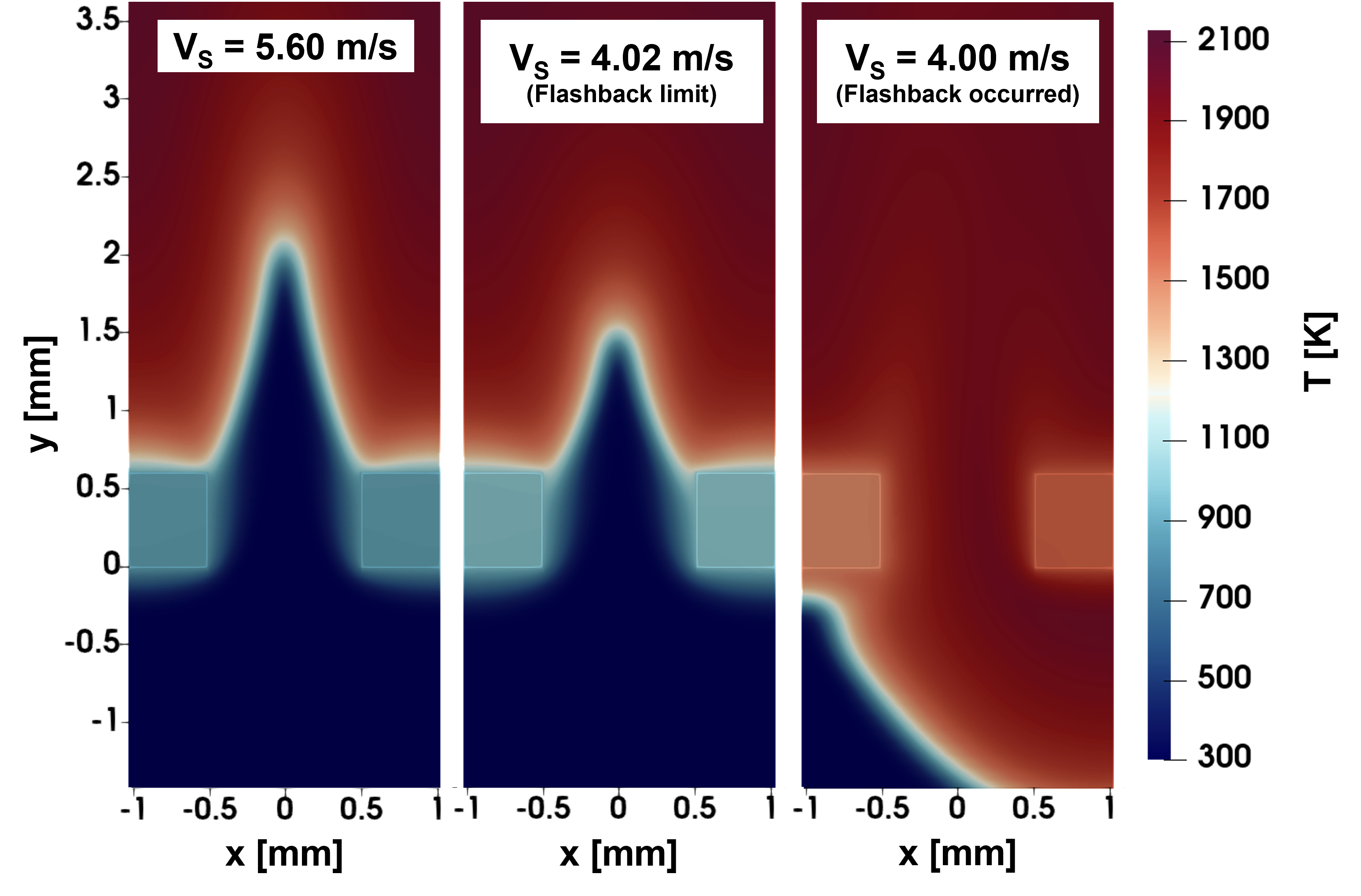}
\caption{Example of the procedure for the estimation of flashback velocity: snapshots of temperature field at decreasing inlet velocity for $\%\ch{H2} = 95\%$, ${\phi = 0.835}$, and ${W = \SI{1}{\mm}}$.}
\label{fig:snapshots}
\end{figure}

Starting from a stable flame characterized by a high inlet velocity, the inlet speed is systematically decreased, using velocity steps, until reaching a critical velocity that induces flashback. The inlet velocity steps used in this procedure are ${\Delta V_{in} = \SI{0.1}{m/s}}$, with an additional refinement of ${\Delta V_{in} = \SI{0.01}{m/s}}$ as the flashback limit is approached. This iterative procedure allows us to estimate the flashback velocity of the investigated mixture. We define the flashback velocity, $V_{FB}$, as the \emph{cold-flow} bulk velocity at the slit entry, defined in Eq.~\ref{eq:defv}, just before flashback occurs:
\begin{equation}
    V_{FB}=\left. V_S \right\vert_{FB} = \frac{D+W}{W} \left. V_{in}\right\vert_{FB}.
\end{equation}
\noindent A representative example of this procedure is illustrated in Fig.~\ref{fig:snapshots}, where we plot the temperature profiles for the case with $\%\ch{H2}=95\%$, ${\phi = 0.835}$ and ${W=\SI{1}{\mm}}$ at decreasing inlet velocities. All fuel percentage values given in this paper are expressed in terms of volume. It can be seen that, as the inlet speed is decreased, the flame approaches the hole entry and the burner temperature increases until a stable configuration is not found anymore and the flame flashes back through the slit. We note that our selection of velocity step sizes for reducing the inlet velocity ensures that this quasi-steady behavior as the flashback velocity is approached is accurately captured across all cases.

To assess the flashback propensity of a specific combination of mixture and geometry parameters, we utilize the flashback velocity scaled by the laminar flame speed of the corresponding mixture, denoted as $V_{FB}/s_L$. The value of $s_L$ is determined for each mixture by analyzing 1D freely propagating flames under the corresponding thermodynamic conditions, employing Cantera~\cite{cantera}.

\section{Stochastic Sensitivity Analysis}\label{sub:SSA}

In this work, a Stochastic Sensitivity Analysis is carried out, using the generalized Polynomial Chaos (gPC) method in its non-intrusive form, to investigate quantitatively how the flashback velocity, $V_{FB}$, its normalized value, $V_{FB}/s_L$, and the volume-averaged burner temperature at the flashback limit, $T_B$, are related to the hydrogen content, $\%\ch{H2}$, the equivalence ratio, $\phi$, and the slit width, $W$. The aim is to obtain continuous response surfaces in the parameter space obtained for each couple of parameters, limiting the number of deterministic simulations and, at the same time, preserving a good level of accuracy of the results. 

In the gPC approach, the dependence between a quantity of interest, $Z$, and the vector of independent uncertain parameters, $\boldsymbol{\zeta}$, is expressed by means of a polynomial expansion~\cite{Ale1}. So, employing term-base indexing:
\begin{equation}
Z(\epsilon) = \displaystyle{\sum_{j=0}^{\infty}} b_j \Psi_j(\boldsymbol{\zeta}(\epsilon))
\label{e:gPC_exp2}
\end{equation}
\noindent where $\epsilon$ is an aleatory event, $\Psi_j(\boldsymbol \zeta)$ is the $j$-th gPC polynomial and $b_j$ is the corresponding projection coefficient. The response surface of the quantity of interest is obtained by truncating the expansion in Eq.~\ref{e:gPC_exp2} to the limit $O$. Applying polynomial order bounds for all one-dimensional polynomials, the truncation limit can be calculated as follows:
\begin{equation}
O= \prod_{k=1}^N(T_k+1)-1
\label{e:gPC_tronc_2}
\end{equation}
\noindent where $N$ is the dimension of the $\boldsymbol{\zeta}$-vector, the index $k$ identifies the particular random variable within that vector, and $T_k$ is the maximum order of the corresponding polynomial. Finally, the coefficient $b_j$ is defined as follows:
\begin{equation}
b_j = \frac{\left< Z, \Psi_j \right>}{\left< \Psi_j, \Psi_j \right>} = \frac{1}{\left< \Psi_j, \Psi_j \right>} \int_{\boldsymbol{\zeta}} Z \Psi_j \omega(\boldsymbol{\zeta}) \text{d} \boldsymbol{\zeta}
\label{e:gPC_coeff}
\end{equation}
\noindent where $\omega(\boldsymbol{\zeta})$ is the weight function connected to $\Psi_j(\boldsymbol{\zeta})$. In this paper, the integral in Eq.~\ref{e:gPC_coeff} is approximated using a Gaussian quadrature formula. The polynomial family $\Psi_j$ has to be \textit{a priori} defined and a suitable choice accelerates the convergence of the procedure. For the Gaussian quadrature, the optimal family is the one having a weight function analogous to the probability measure of the random variables. The choice of the polynomial family thus is related to the shape of the PDF of the uncertain parameters. In this case, since the input random variables are characterized by a uniform PDF, Legendre polynomials are selected. It is expected that, for a given variation interval of the input parameters, the uniform PDF distribution should give the largest variability of the output quantities, thus providing a ‘conservative’ estimation of the sensitivity to the considered input parameters. The polynomial expansion is truncated to the $3^{rd}$ order in each dimension and thus, 4 quadrature points are necessary for each variable (Gauss-Legendre points).

It is worth noting that the UQ analysis could be performed by considering the three parameters, i.e., \ch{H2} content, equivalence ratio $\phi$, and slit width $W$, all together, resulting in 64 simulation points. However, we decided to carry out three different stochastic sensitivity analyses for each couple of parameters keeping the third one fixed. This leads to 16 simulations for each analysis and, thus, to a total of 48. The gain in computational effort is not negligible, as each simulation point implies a number of transient numerical simulations to observe flashback. Moreover, by considering the 3 parameters together, for some combinations of the values, flashback does not occur, and this would introduce possible discontinuities in the response surfaces that cannot be analyzed by the gPC approach. Conversely, by varying only two parameters, this can be avoided by ad-hoc tailoring the ranges of variation. 
Indeed, in each analysis, the ranges of the considered parameters are selected to encompass the broadest possible region within the parameter space where flashback phenomena can occur. For the \%\ch{H2}/$\phi$ analysis, we use ${\%\ch{H2}\in [85\%,100\%]}$ and ${\phi \in [0.5,1.0]}$, with ${W=\SI{1}{\mm}}$. For the \%\ch{H2}/$W$ analysis, we use ${\%\ch{H2}\in [80\%,100\%]}$ and ${W \in [\SI{0.5}{\mm},\SI{1.2}{\mm}]}$, with $\phi=0.7$. Finally, for the $\phi$/$W$ analysis, we use ${\phi \in [0.5,1.0]}$ and ${W \in [\SI{0.5}{\mm},\SI{1.2}{\mm}]}$, with $\%\ch{H2} = 100\%$. Outside of this region, specifically when reducing the values of $\phi$, \ch{H2} content, or $W$ while keeping the remaining parameters constant, a decrease in input velocity leads to flame quenching instead of flashback. Consequently, it becomes impossible to define a flashback velocity in such cases. The quadrature points, i.e. the values of the parameters at which the computations are carried out, are reported in Tab.~\ref{tab:quadrature_points_H2_phi},~\ref{tab:quadrature_points_H2_W}, and~\ref{tab:quadrature_points_phi_W} for each considered analysis.

In addition, two independent simulations are performed for each analysis to test the reliability of the response surfaces. These test points are chosen a posteriori to be distant from each other and from the quadrature points, placed in opposite regions of the parameter space. The variability of the output quantities is described in terms of total variance as $\sigma^2 = \sum_{j=1} (b^*_j)^2$, with $b^*_j = b_j |\Psi_j|$, $|\Psi_j|$ being the norm of the $j$-th polynomial. The sensitivity of the quantities of interest to a single input parameter or to a combination of them is computed using the variance decomposition method proposed by~\cite{SOBOL2001271}. The Sobol' index (also called sensitivity index) $I_i$ is defined as the ratio between the partial variance $\sigma^2_i$, i.e., the variance only due to the $i-$th uncertain input parameter, and the total variance $\sigma^2$, as $I_i = \sigma^2_i/\sigma^2$. In the case of a two-parameter analysis, the Sobol' index relative to the interaction between the two parameters is equal to $I_{1,2}=1-I_1-I_2$.

\begin{table}[!ht]
    \centering
    \begin{tabular}{ccccc}
        Quadrature Points & $1^{st}$ & $2^{nd}$ & $3^{rd}$ & $4^{th}$\\\hline
        \ch{H2} content [\%] & 86.0 & 89.9 & 95.0 & 98.9\\
        $\phi$ [-] & 0.535  & 0.665 & 0.835 & 0.965\\\hline
    \end{tabular}
    \caption{Quadrature points for the two input parameters: \ch{H2} content and equivalence ratio.}
    \label{tab:quadrature_points_H2_phi}
\end{table}

\begin{table}[!ht]
    \centering
    \begin{tabular}{ccccc}
        Quadrature Points & $1^{st}$ & $2^{nd}$ & $3^{rd}$ & $4^{th}$\\\hline
        \ch{H2} content [\%] & 81.4  &  86.6 &  93.4 & 98.6 \\
        $W$ [$\SI{}{\mm}$] & 0.55 & 0.73 & 0.97 & 1.15 \\ \hline
    \end{tabular}
    \caption{Quadrature points for the two input parameters: \ch{H2} content and slit width.}
    \label{tab:quadrature_points_H2_W}
\end{table}

\begin{table}[!ht]
    \centering
    \begin{tabular}{ccccc}
        Quadrature Points & $1^{st}$ & $2^{nd}$ & $3^{rd}$ & $4^{th}$\\\hline
        $\phi$ [-] & 0.535  & 0.665 & 0.835 & 0.965\\
        $W$ [$\SI{}{\mm}$] & 0.55 & 0.73 & 0.97 & 1.15\\\hline
    \end{tabular}
    \caption{Quadrature points for the two input parameters: equivalence ratio and slit width.}
    \label{tab:quadrature_points_phi_W}
\end{table}

\section{Results}

\subsection{Effect of \ch{H2} content and equivalence ratio}\label{sub:h2phi}

The sensitivity analysis of the flashback velocity, $V_{FB}$, its normalized value, $V_{FB}/s_L$, and the burner plate temperature at the flashback limit, $T_B$, is performed as the \ch{H2} content and the equivalence ratio vary in the ranges ${\%\ch{H2} \in [85\%,100\%]}$ and ${\phi \in [0.5,1.0]}$. The slit width is fixed at ${W=\SI{1}{\mm}}$, which is chosen to ensure a sufficiently large area in the parameter space for defining the flashback velocity. It is important to emphasize that when lowering $\phi$ while keeping the \ch{H2} content at 85\%, or when reducing the \ch{H2} content while maintaining $\phi=0.5$, we observe flame quenching for low inlet velocity, making it impossible to define a flashback velocity in those cases. After obtaining the response surfaces, two test simulations are conducted for the points with $\%H2 = 87.5\%$ and $\phi = 0.9$, as well as $\%H2 = 97.5\%$ and $\phi = 0.6$. The results for both $V_{FB}$ and $T_B$ are found to be consistent with the response surface values at these points, with an error of less than 2\%.

\paragraph{Flashback velocity}

\begin{figure}[!ht]
    \centering
    \subfigure[]{\includegraphics[width=0.55\textwidth]{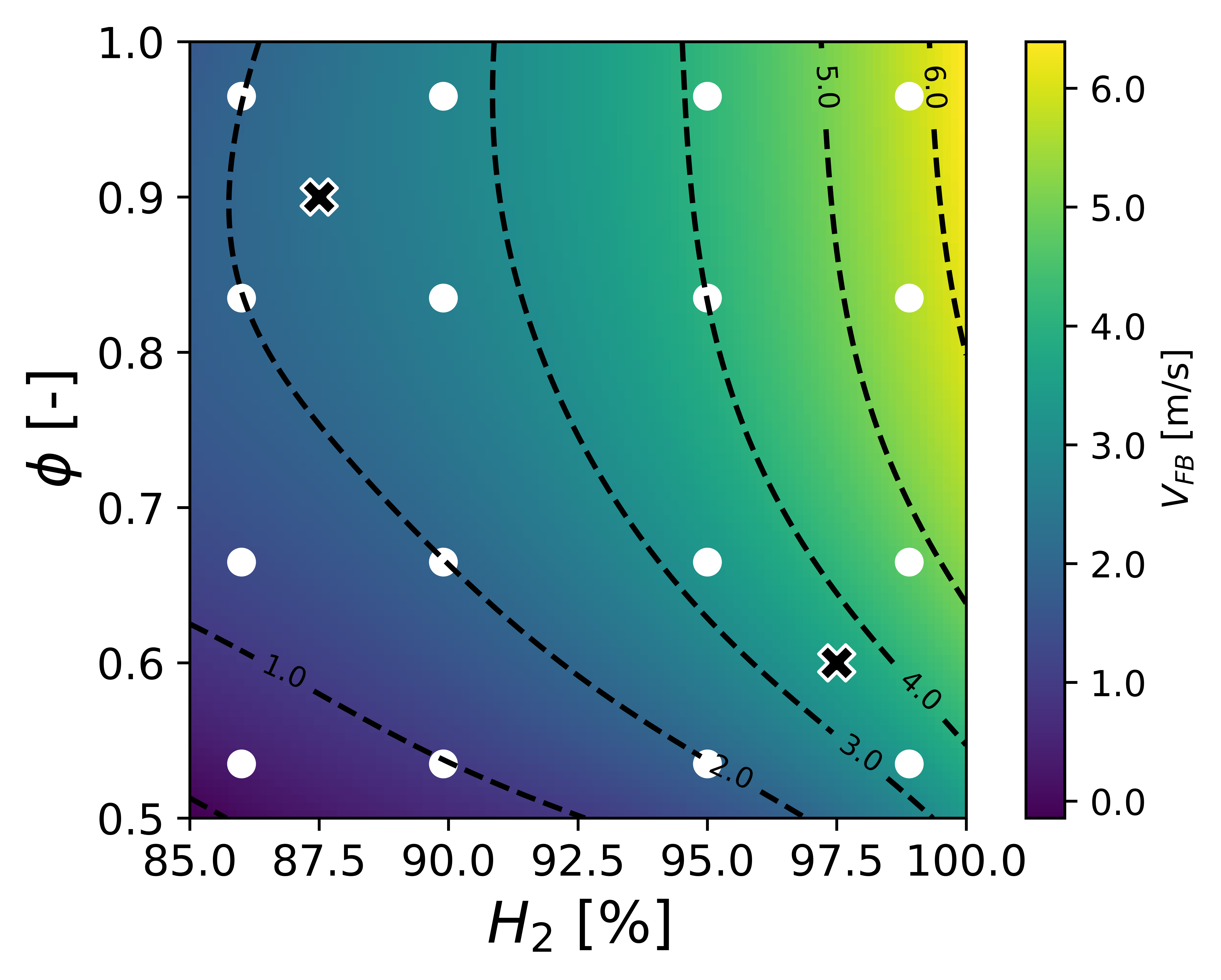}}
    \subfigure[]{\includegraphics[width=0.35\textwidth]{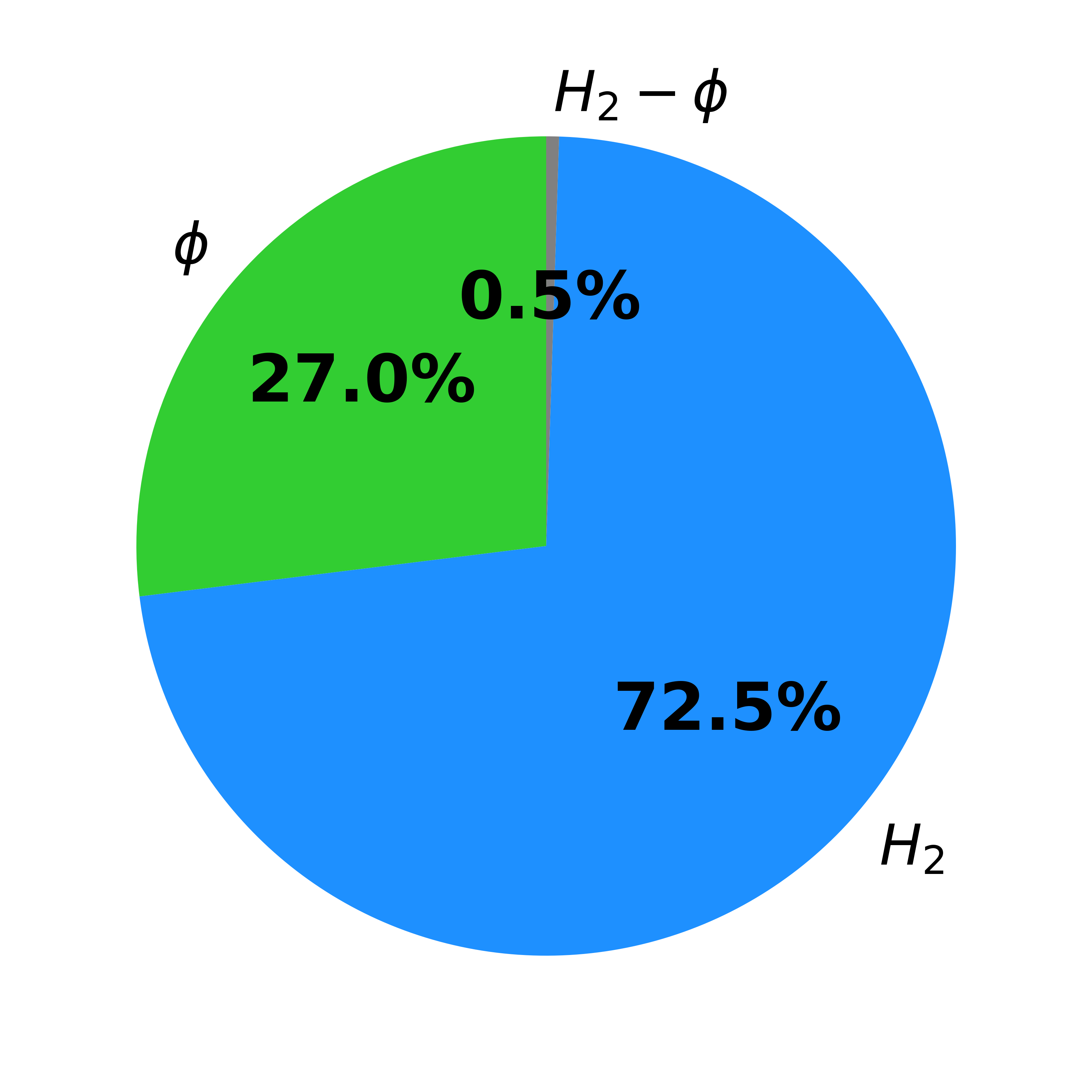}}
    \caption{(a) Stochastic response surface of $V_{FB}$ in the $\%\ch{H2}$ - $\phi$ parameter space. White dots represent quadrature points. Black crosses indicate test points.} (b) Sobol' indices.
    \label{fig:H2_phi_VF}
\end{figure}

Fig.~\ref{fig:H2_phi_VF} shows the flashback velocity $V_{FB}$ as a colored map in the \ch{H2} content - $\phi$ space, with the relative Sobol' indices given as percentages. 
\noindent We observe that $V_{FB}$ ranges from \SI{0.2}{m/s}, for low \%\ch{H2} and low $\phi$, to \SI{6}{m/s}, for high \%\ch{H2} and high $\phi$. The dominant input parameter influencing $V_{FB}$ is the hydrogen content, with a Sobol' index of $I_{\ch{H2}}=72.6\%$. There is also a weaker but still significant dependence on the equivalence ratio ($\phi$), with a Sobol' index of $I_{\phi}=26.9\%$. Interestingly, the sensitivity of $V_{FB}$ to these two parameters is consistent across the entire parameter space, as indicated by the interaction Sobol' index $I_{\ch{H2},\phi}$ being close to zero. 

The trend of $V_{FB}$ in the parameter space, which shows an increase in flashback velocity with increasing hydrogen content and equivalence ratio independently, mimics the well-known trend of the laminar flame speed $s_L$. To better understand the flashback physics, it is more appropriate to consider the  $V_{FB}/s_L$ ratio. Indeed, this normalization eliminates the trivial dependence on the variability of $s_L$ and allows for a clearer assessment of the differences in the flashback phenomena.

\paragraph{Normalized flashback velocity}

\begin{figure}[!ht]
    \centering
    \subfigure[]{\includegraphics[width=0.55\textwidth]{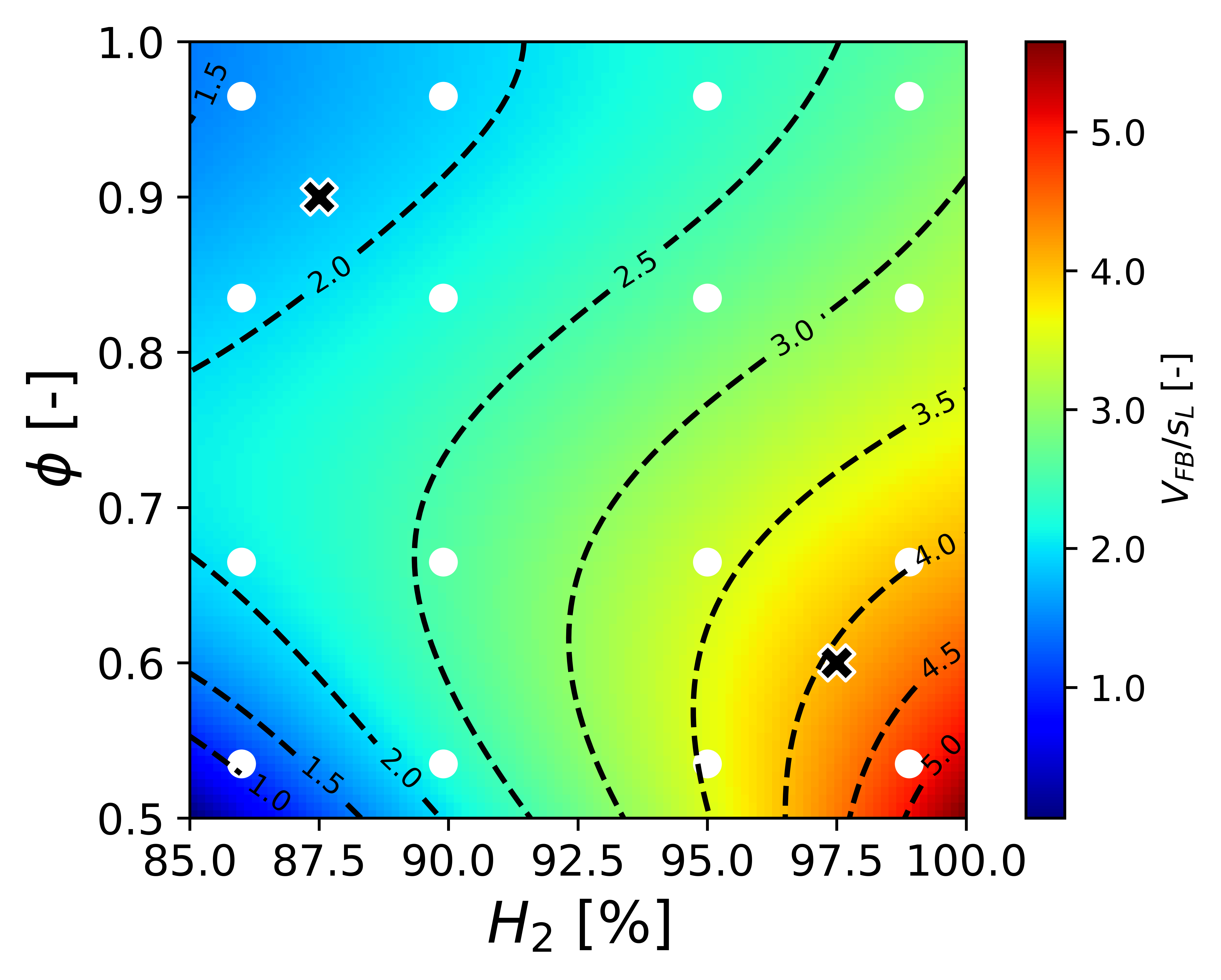}}
    \subfigure[]{\includegraphics[width=0.35\textwidth]{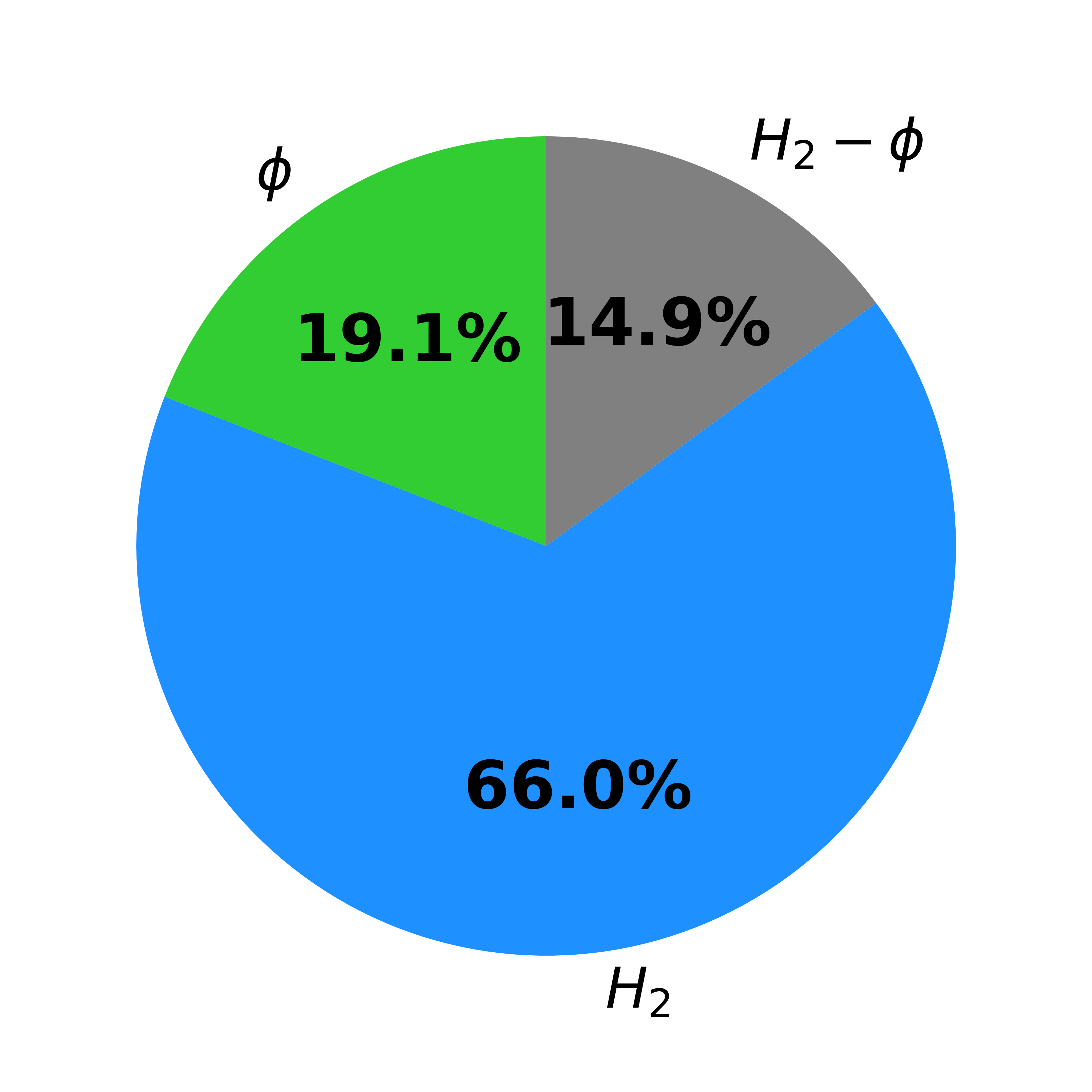}}
    \caption{(a) Stochastic response surface of $V_{FB}/s_L$ in the $\%\ch{H2}$ - $\phi$ parameter space. White dots represent quadrature points. Black crosses indicate test points.} (b) Sobol' indices.
    \label{fig:H2_phi_Vscaled}
\end{figure}

Fig.~\ref{fig:H2_phi_Vscaled} shows the $V_{FB}/s_L$ map in the parameter space and the relative Sobol' indices.
\noindent The variation of the normalized flashback velocity over the parameter space is significant, ranging from $V_{FB}/s_L \lesssim 1.0$ to $V_{FB}/s_L \gtrsim 5.0$. The minimum is observed in the region with low \ch{H2} content and low equivalence ratio, very close to the quenching limit, probably due to the low reactivity of the mixture. The maximum is observed in the region with high \ch{H2} content and low equivalence ratio. The hydrogen content is the most influential parameter, with $I_{\ch{H2}}$ being 66.0\%, while the equivalence ratio contributes to 19.1\% of the variability. Moreover, as confirmed by the interaction Sobol' index $I_{\ch{H2},\phi}=14.9\%$, we observe a larger impact of $\phi$ for high \ch{H2} contents, with the maximum $V_{FB}/s_L$ occurring in the zone with the highest \ch{H2} content and the lowest equivalence ratio. 

\begin{figure}[!ht]
    \centering
    \subfigure[]{\includegraphics[width=0.48\textwidth]{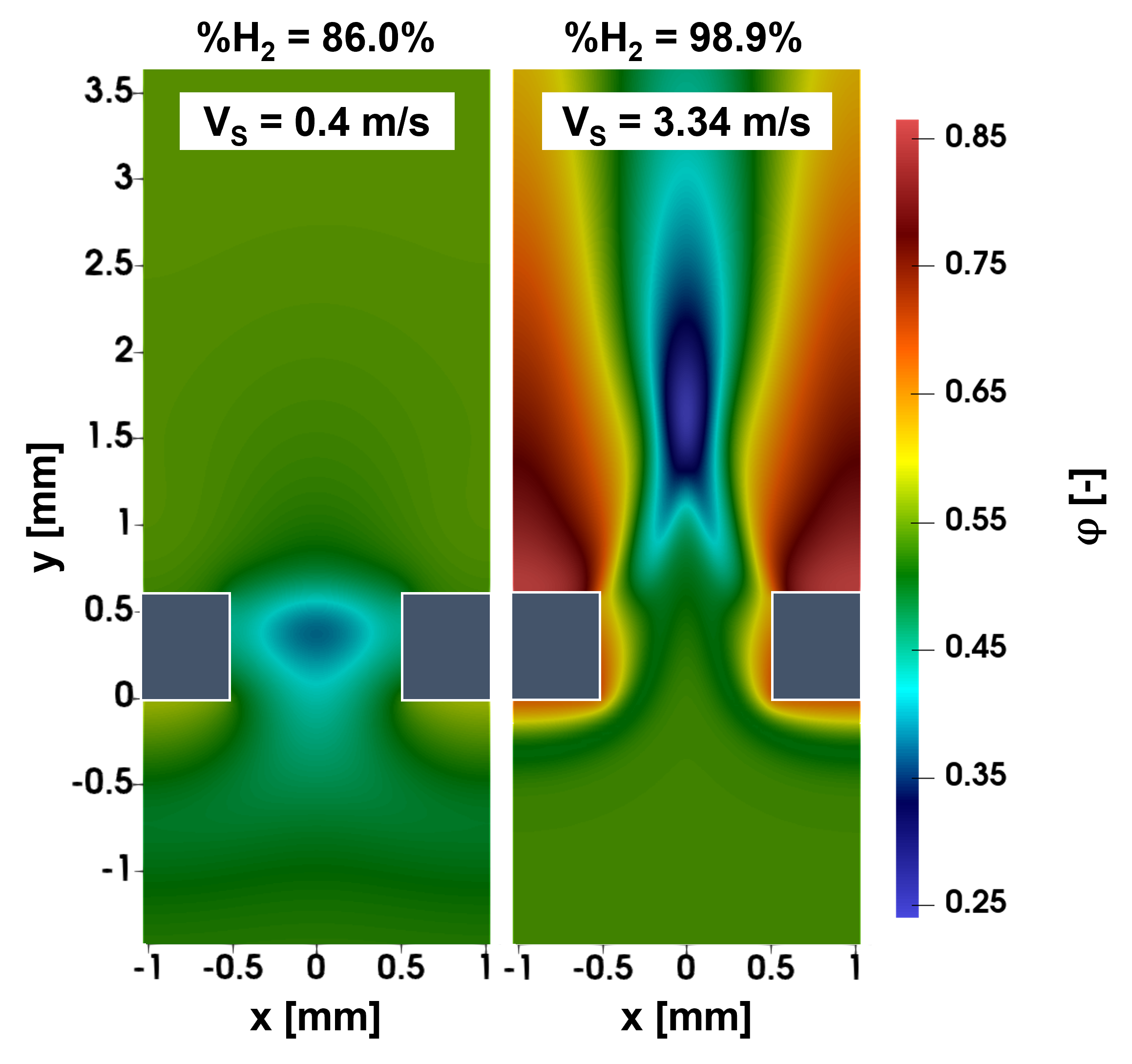}}
    \subfigure[]{\includegraphics[width=0.48\textwidth]{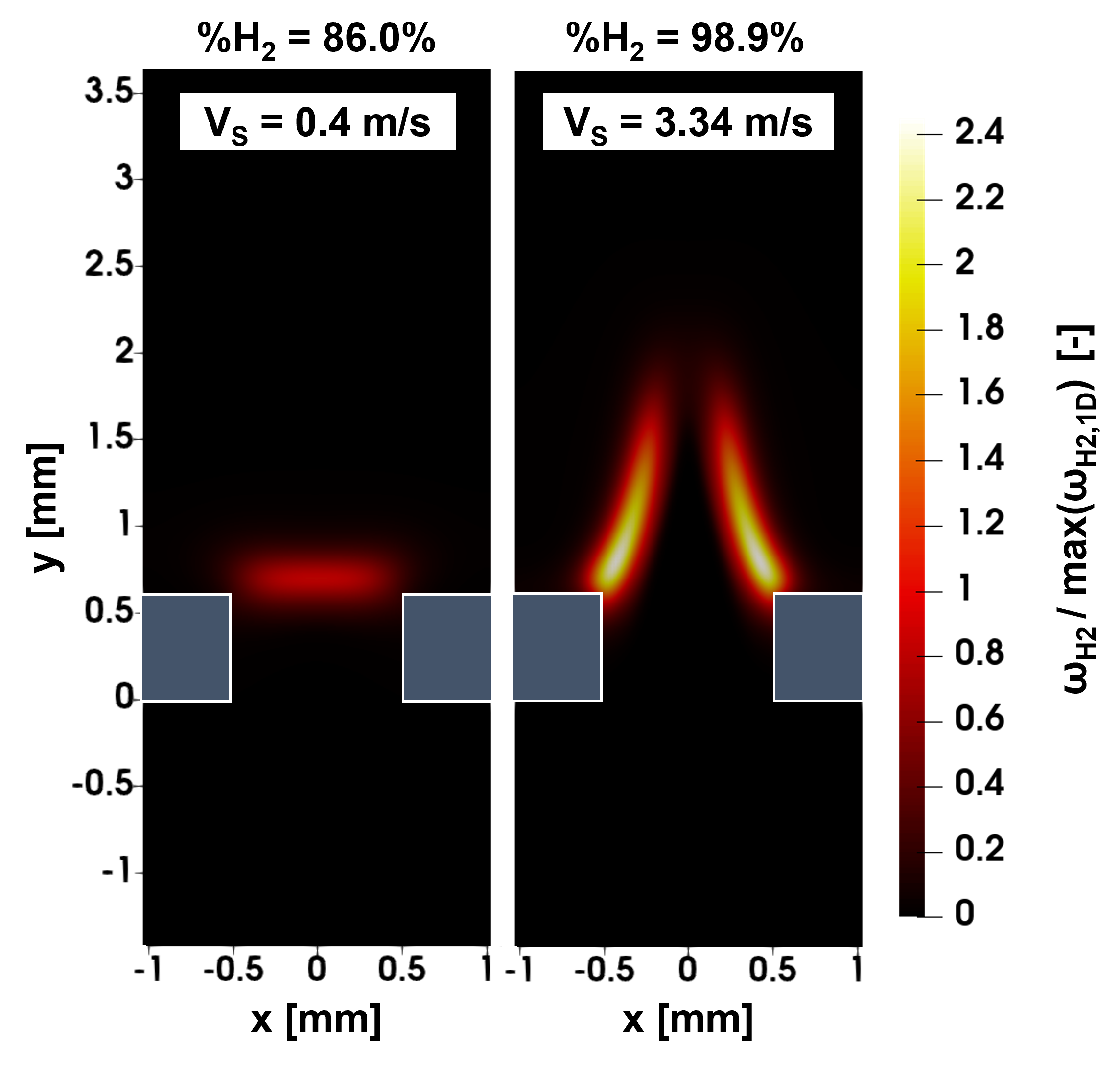}}
    \caption{(a) Local equivalence ratio for flashback limit cases with $\%\ch{H2}=86\%$ and $\%\ch{H2}=98.9\%$ at $\phi=0.535$. (b) Normalized molecular hydrogen consumption rate for flashback limit cases with $\%\ch{H2}=86\%$ and $\%\ch{H2}=98.9\%$ at $\phi=0.535$.}
    \label{fig:H2_phi_contours}
\end{figure}

\begin{figure}[!ht]
\centering
\includegraphics[width=0.5\textwidth]{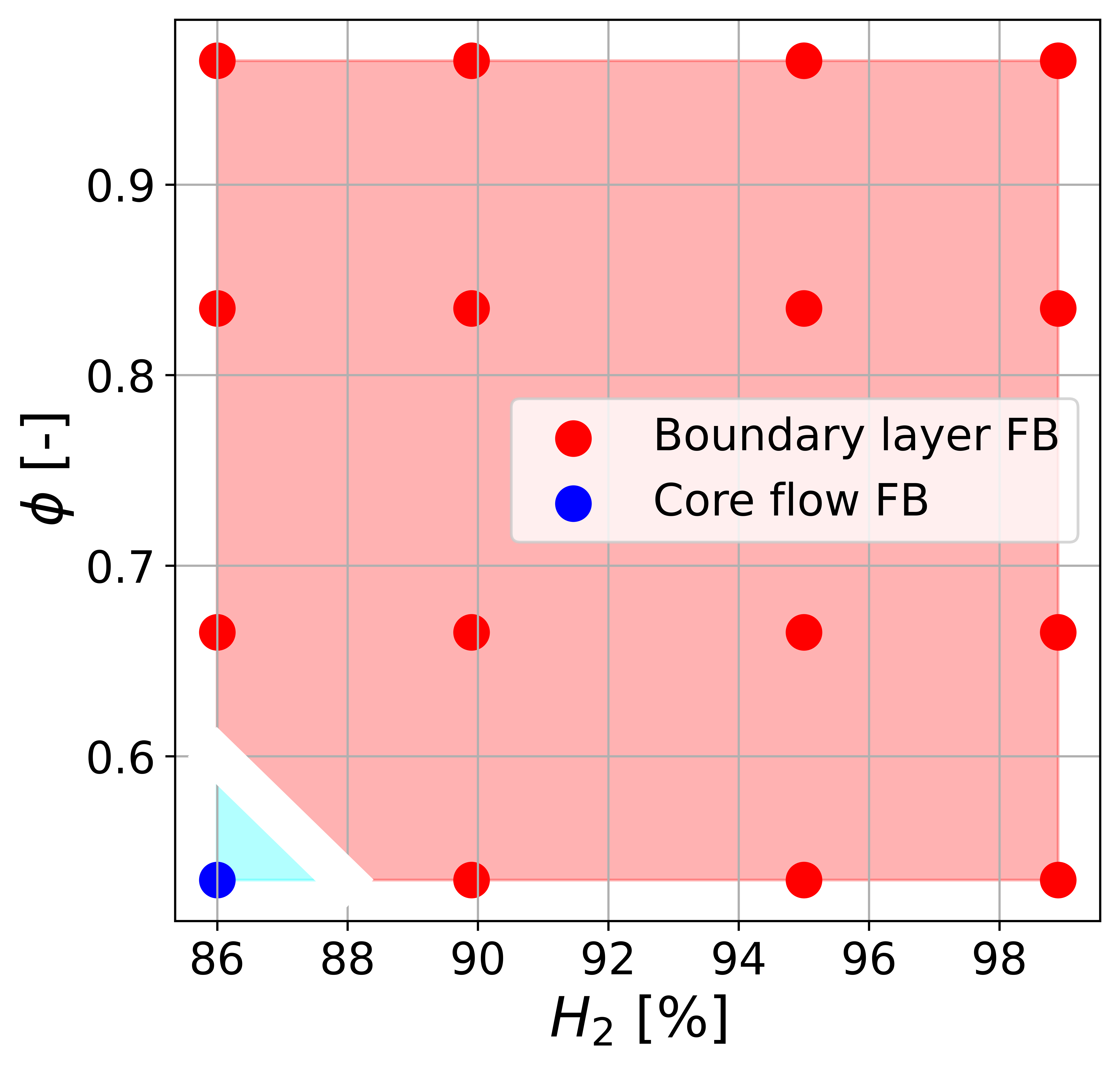}
\caption{Flashback regimes in the $\%\ch{H2}$ - $\phi$ parameter space. Blue represents ``core flow flashback", and red represents ``boundary layer flashback".}
\label{fig:border_H2_phi}
\end{figure}

As previously mentioned, the ratio $V_{FB}/s_L$ is the most effective parameter for assessing flashback propensity, as it highlights the differences in the flashback phenomena, eliminating the trivial dependence on the value of the laminar flame speed. To visualize and better understand the underlying physics, Fig.~\ref{fig:H2_phi_contours} presents the local equivalence ratio, $\varphi$, and the normalized molecular \ch{H2} consumption rate, $\omega_{\ch{H2}}/\textnormal{max}(\omega_{\ch{H2},\textnormal{1D}})$, for the flashback limits of the cases with $\%\ch{H2}=86.0\%$ and $\%\ch{H2}=98.9\%$ at $\phi=0.535$. $\textnormal{max}(\omega_{\ch{H2},\textnormal{1D}})$ is the maximum \ch{H2} consumption rate obtained in the corresponding 1D flame. Following Pope et al.~\cite{POPE1999605}, the local equivalence ratio $\varphi$ is defined by:
\begin{equation}
    \varphi=\frac{2\ \chi_C+1/2\ \chi_H}{\chi_O};\hspace{10pt}\chi_l=\sum_{k=1}^{N_{\textnormal{sp}}}a_{k,l}X_k,
\end{equation}
where $\chi_l$ denotes the mole fraction of element $l$, $X_k$ is the mole fraction of species $k$, and $a_k,l$ is the number of atoms of element $l$ contained in a molecule of species $k$. It is important to note that, since the unburnt mixture is homogeneous, variations in the local equivalence ratio can only arise due to preferential diffusion. The snapshots are obtained from simulations performed at the corresponding quadrature points and depict the flashback limit, i.e., the last stable flame before the occurrence of flashback.
\noindent We observe that the hydrogen content has a significant influence on the effect of preferential diffusion on the local equivalence ratio, leading to a local enrichment of up to $\varphi=0.86$ in the flame base zone for the case with 98.9\% \ch{H2}. In addition to preferential diffusion effects caused by curvature, typically occurring in mixtures with $Le<1$, we observe Soret-induced preferential diffusion effects. As shown by Vance et al.~\cite{VANCEsoret}, since the Soret coefficients of light species like \ch{H2} and \ch{CH4} are negative (diffusion from cold to hot regions), while heavier species like \ch{O2} have positive Soret coefficients (diffusion from hot to cold regions), a higher $\varphi$ is observed in close proximity to the hot surfaces, where temperature gradients are high. The local enrichment induced by both sources of preferential diffusion intensifies the flame in the flame base region, resulting in an increased flame speed in the presence of high velocity gradients near the wall. This effect favors flashback by modifying $s_L$ in the critical region, thus leading to a higher value of $V_{FB}/s_L$. In such cases, we observe flashback occurring asymmetrically at the burner wall. For the cases with lower hydrogen (\ch{H2}) content, preferential diffusion effects do not play a significant role. When the inlet velocity is decreased, flashback occurs smoothly in this case, with the flame front moving towards the slit entry, flattening, heating up the burner, and then passing through the slit symmetrically. We refer to these two distinct flashback regimes as the ``boundary layer flashback" regime and the ``core flow flashback" regime, respectively~\cite{FRUZZA2023}. In Fig.~\ref{fig:border_H2_phi}, we partition the parameter space into two regions corresponding to the two flashback regimes based on observations derived from transient simulations conducted at the quadrature points. It can be observed that, for the selected parameter set, the prevalence of the ``boundary layer flashback" is evident, with this phenomenon occurring at all quadrature points except for the one characterized by lower $\%\ch{H2}$ and $\phi$, where ``core flow flashback" is observed instead. It is worth noting that the transition between these two flashback regimes corresponds to values of $V_{FB}/s_L$ between 1 and 2, as already observed in Fruzza et al.~\cite{FRUZZA2023}.

The role of preferential diffusion in promoting flashback explains why the higher values of $V_{FB}/s_L$ are observed in the parameter space region characterized by the highest $\%\ch{H2}$ and the lowest $\phi$, where the influence of preferential diffusion effects is expected to be more pronounced.

\paragraph{Burner temperature}

\begin{figure}[!ht]
    \centering
    \subfigure[]{\includegraphics[width=0.55\textwidth]{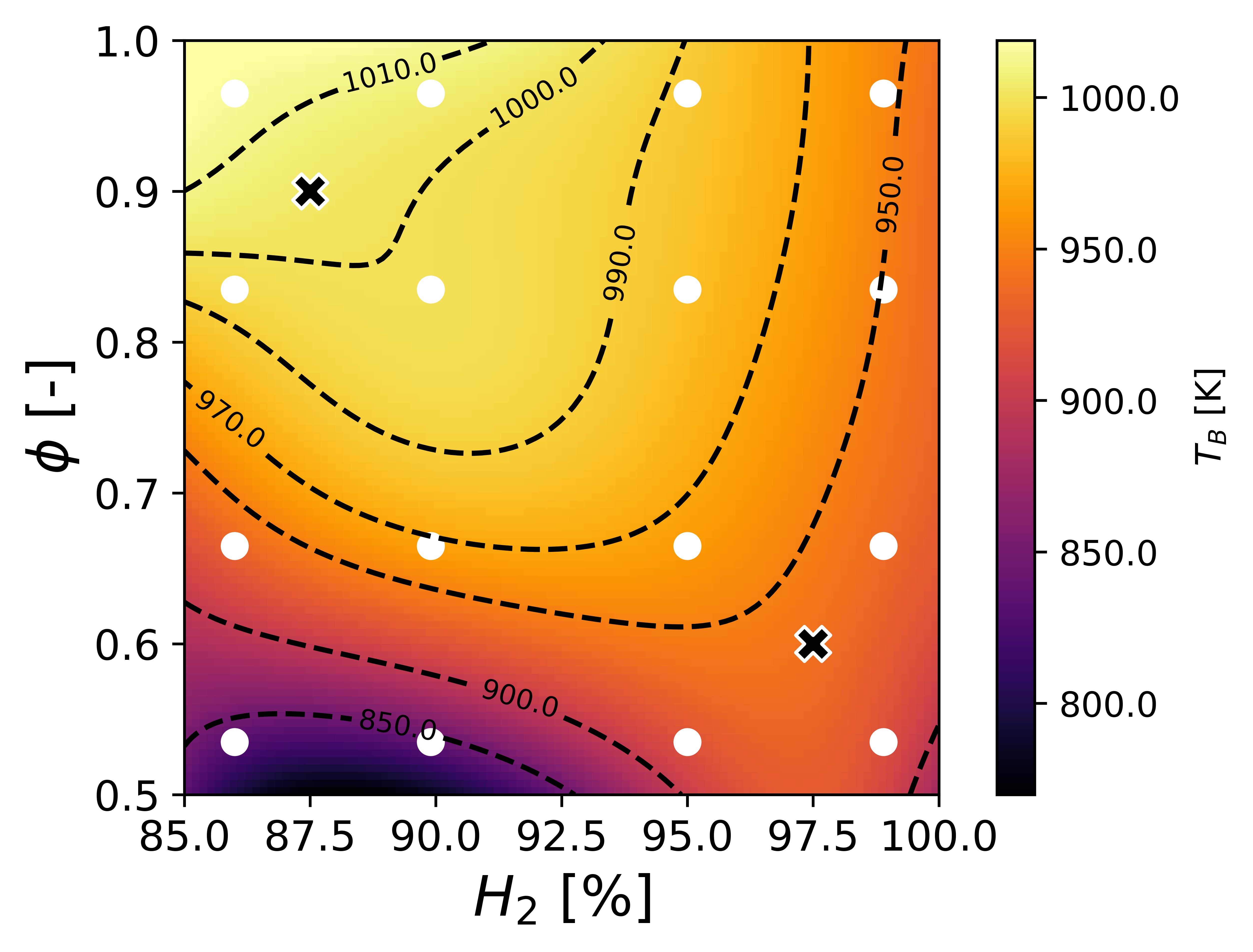}}
    \subfigure[]{\includegraphics[width=0.35\textwidth]{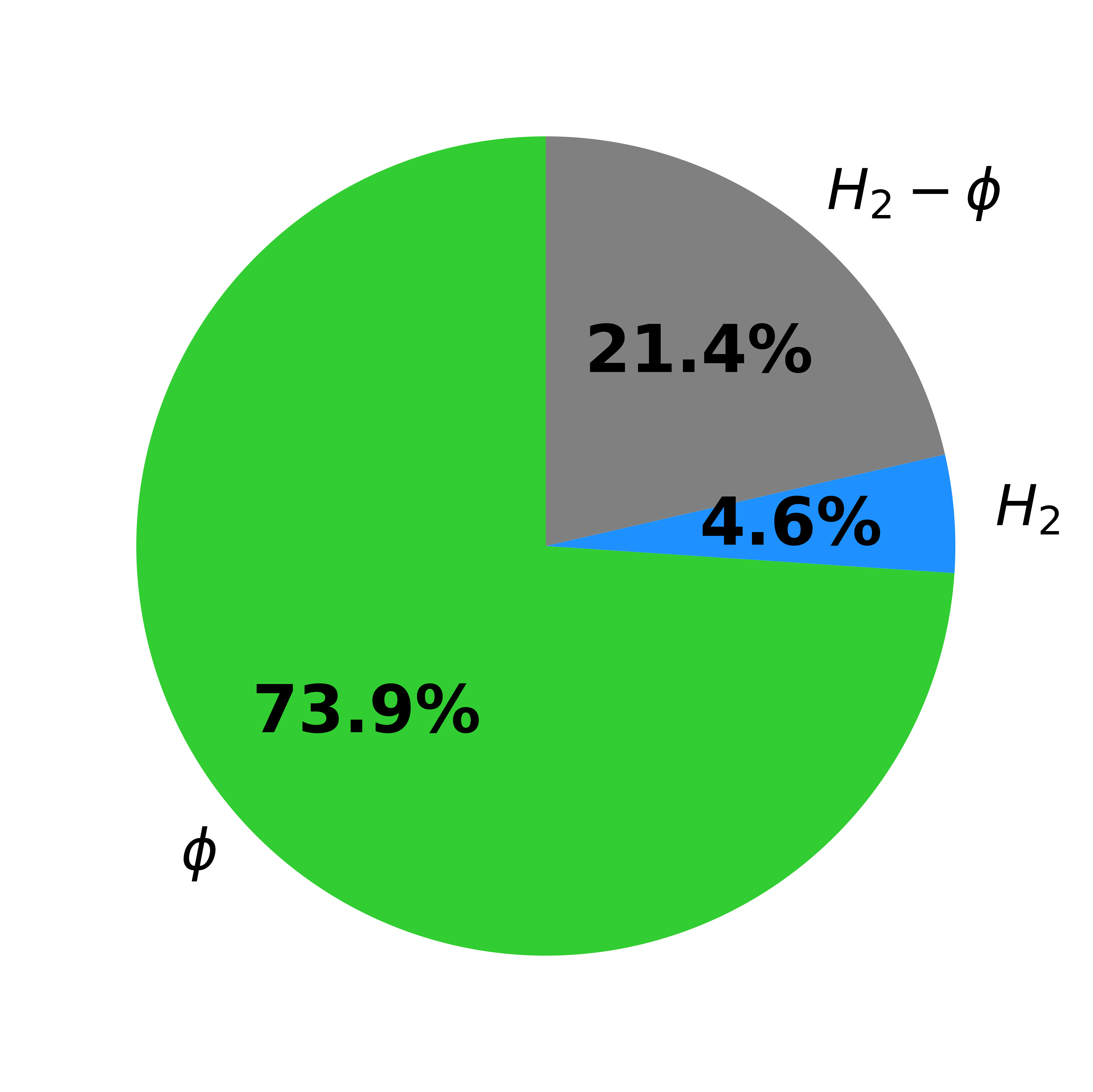}}
    \caption{(a) Stochastic response surface of $T_B$ in the $\%\ch{H2}$ - $\phi$ parameter space. White dots represent quadrature points. Black crosses indicate test points.} (b) Sobol' indices.
    \label{fig:H2_phi_T}
\end{figure}

Fig.~\ref{fig:H2_phi_T} shows the burner plate temperature at the flashback limit, which represents the maximum attainable temperature of the burner, in the parameter space, and the relative Sobol' indices.
\noindent We observe a significant variation in the maximum temperature of the burner plate within the investigated parameter space. The minimum, $T_B \approx \SI{825}{\K}$, is observed for low \ch{H2} content and low equivalence ratio, while the maximum, $T_B \approx \SI{1010}{\K}$, is observed for low \ch{H2} content and high equivalence ratio. In most of the parameter space, the variability of $T_B$ is primarily attributed to changes in $\phi$. This is supported by the Sobol' indices, $I_\phi=74.0\%$ and $I_{\ch{H2}}=4.6\%$. However, as the hydrogen content approaches 100\%, $T_B$ becomes almost independent of $\phi$, with $T_B \approx \SI{950}{\K}$ for all equivalence ratios. The difference in the sensitivity of $T_B$ to $\phi$ in different regions of the parameter space is quantified by the interaction Sobol' index $I_{\ch{H2},\phi}=21.4\%$.

The variability of $T_B$ in the parameter space can be attributed to various factors, including the position of the flame at the flashback limit, the total mass flow rate, radiative losses, the flame attachment to the burner plate, and the contact area between the hot burnt gases and the top surface of the burner plate. Due to the complexity of heat exchange phenomena between the flame, burner, and external environment, understanding and predicting the interplay of these effects is challenging. Therefore, the construction of a map in the parameter space using the gPC and the Stochastic Sensitivity Analysis provide a valuable approach for qualitatively predicting the maximum burner temperature when varying the parameters of interest.

\subsection{Effect of \ch{H2} content and slit width}\label{sub:h2w}

A similar analysis is conducted by varying the \ch{H2} content in the range ${\%\ch{H2} \in [80\%,100\%]}$ and the slit width in the range ${W \in [\SI{0.5}{\mm},\SI{1.2}{\mm}]}$, while keeping a constant equivalence ratio of $\phi=0.7$. This specific value of $\phi$ is chosen to enable the definition of a flashback velocity throughout the entire parameter space, while ensuring the mixtures remain sufficiently lean, aligning with practical considerations. Two test simulations are carried out at the points with $\%H2 = 83\%$ and $W = \SI{0.65}{\mm}$, and $\%H2 = 97\%$ and $W = \SI{1.05}{\mm}$: the results for both $V_{FB}$ and $T_B$ are found to be consistent with the response surface with an error of less than 1\%.

\paragraph{Flashback velocity}

\begin{figure}[!ht]
    \centering
    \subfigure[]{\includegraphics[width=0.55\textwidth]{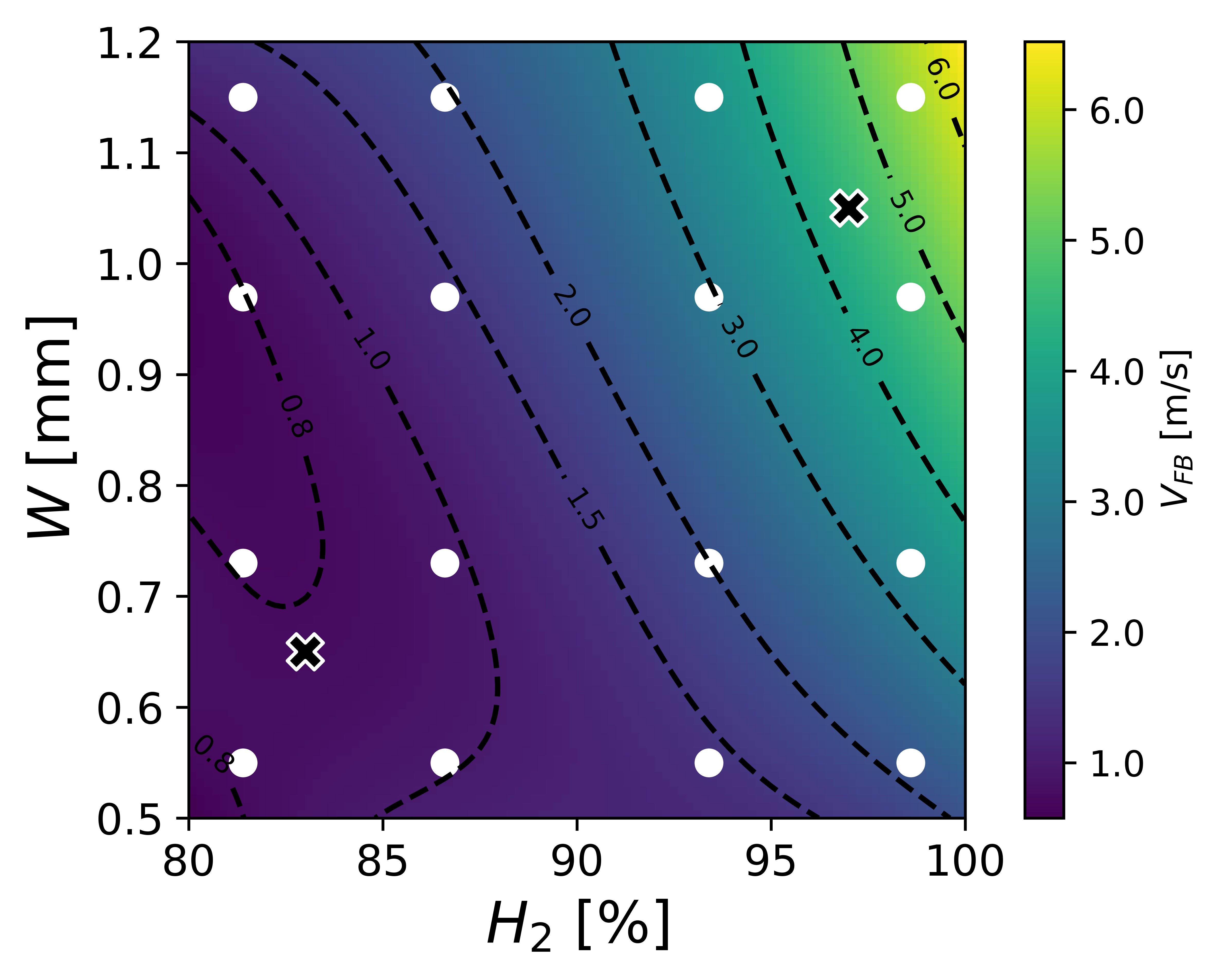}}
    \subfigure[]{\includegraphics[width=0.35\textwidth]{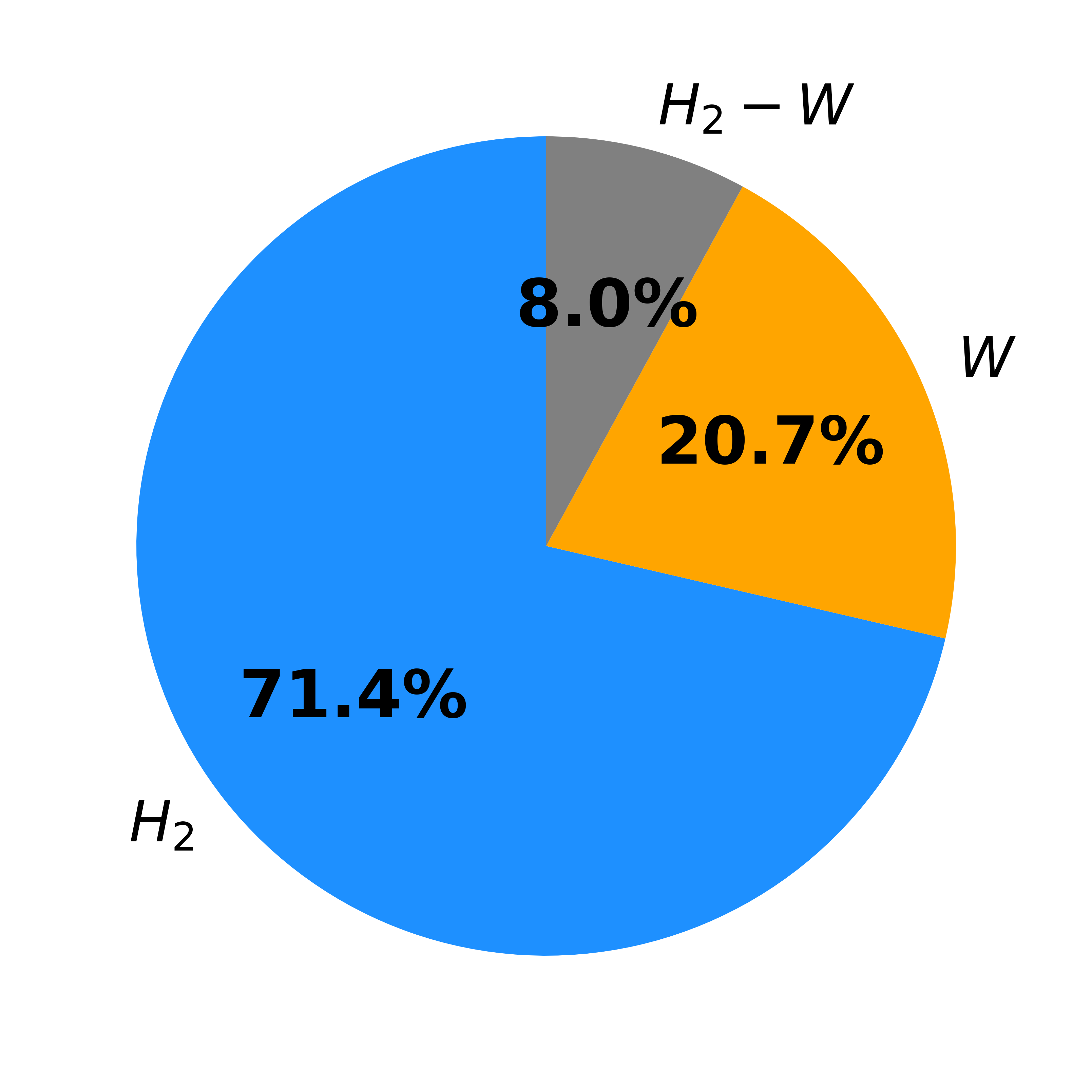}}
    \caption{(a) Stochastic response surface of $V_{FB}$ in the $\%\ch{H2}$ - $W$ parameter space. White dots represent quadrature points. Black crosses indicate test points.} (b) Sobol' indices.
    \label{fig:H2_W_V}
\end{figure}

Fig.~\ref{fig:H2_W_V} illustrates the relationship between the flashback velocity and the two input parameters, as well as the corresponding Sobol' indices.
\noindent The flashback velocity demonstrates significant variability across the entire parameter space, with the maximum value, $V_{FB,\textnormal{max}}\simeq\SI{6.1}{m/s}$, being nearly ten times larger than the minimum, $V_{FB,\textnormal{min}}\simeq\SI{0.7}{m/s}$. The variation of $V_{FB}$ is primarily influenced by the \ch{H2} content, with a Sobol' index of $I_{\ch{H2}}=71.4\%$. However, the slit width also exhibits a non-negligible impact on $V_{FB}$, being $I_{W}=20.7\%$, but only for high \ch{H2} contents, at which we observe the flashback velocity ranging from $V_{FB}\simeq\SI{2.3}{m/s}$ to $V_{FB}\simeq\SI{6.1}{m/s}$.

The underlying physical reasons for the non-negligible dependence of $V_{FB}$ on the $W$ are not straightforward. Once again, analyzing the variation of the normalized value $V_{FB}/s_L$ provides a better approach to understanding the involved phenomena.

\paragraph{Normalized flashback velocity}

\begin{figure}[!ht]
    \centering
    \subfigure[]{\includegraphics[width=0.55\textwidth]{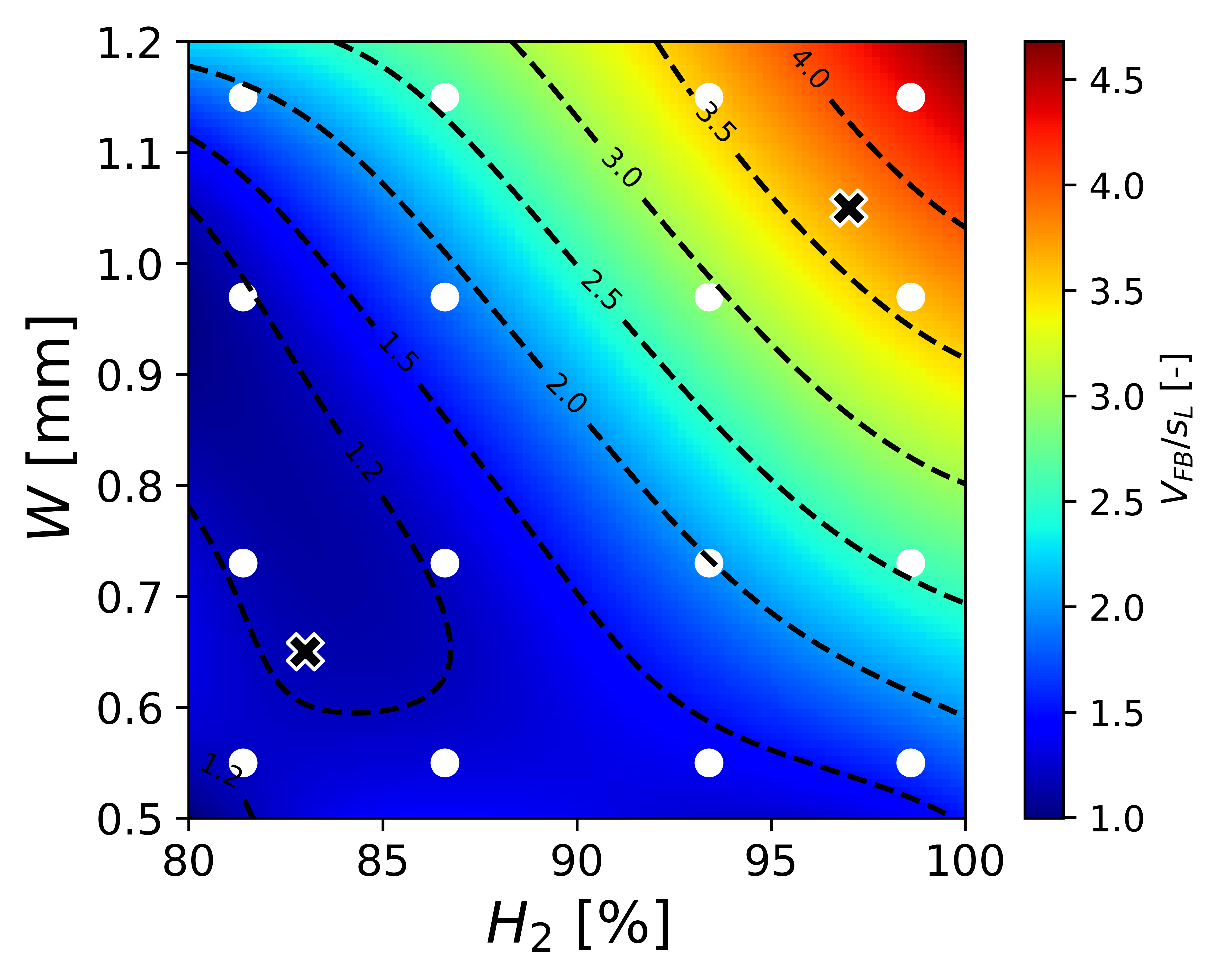}}
    \subfigure[]{\includegraphics[width=0.35\textwidth]{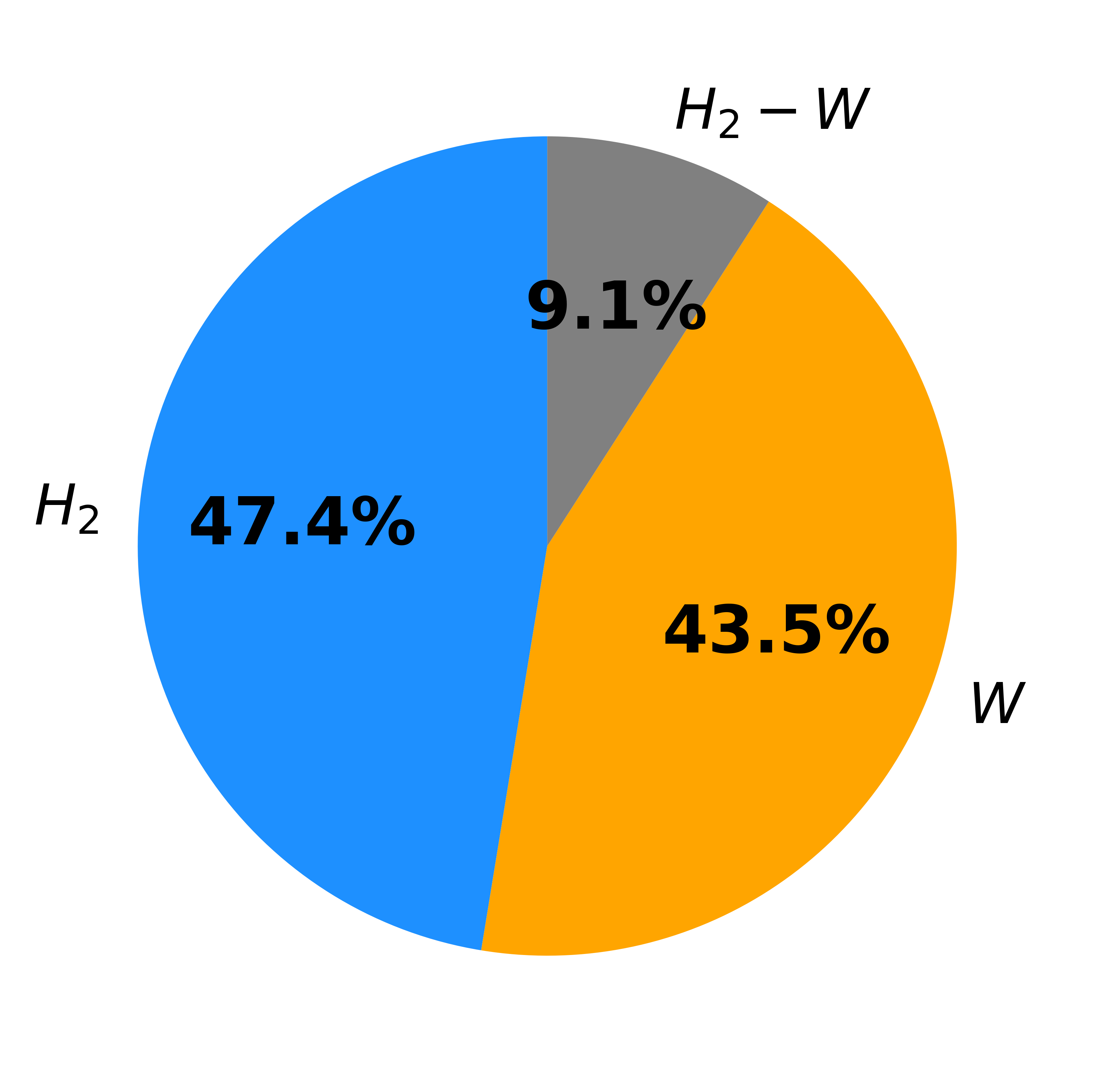}}
    \caption{(a) Stochastic response surface of $V_{FB}/s_L$ in the $\%\ch{H2}$ - $W$ parameter space. White dots represent quadrature points. Black crosses indicate test points.} (b) Sobol' indices.
    \label{fig:H2_W_Vscaled}
\end{figure}

Fig.~\ref{fig:H2_W_Vscaled} illustrates $V_{FB}/s_L$ in the parameter space and the corresponding Sobol' indices.
\noindent The scaled flashback velocity exhibits a significant variability of approximately 450\% across the parameter space, ranging from 1 to 4.5. Interestingly, the dependence of $V_{FB}/s_L$ on $W$ is equally important as the dependence on $\%\ch{H2}$, as indicated by the similar values of their respective Sobol' indices, $I_{W}=43.5\%$ and $I_{\ch{H2}}=47.4\%$.

\begin{figure}[!ht]
    \centering
    \subfigure[]{\includegraphics[width=0.48\textwidth]{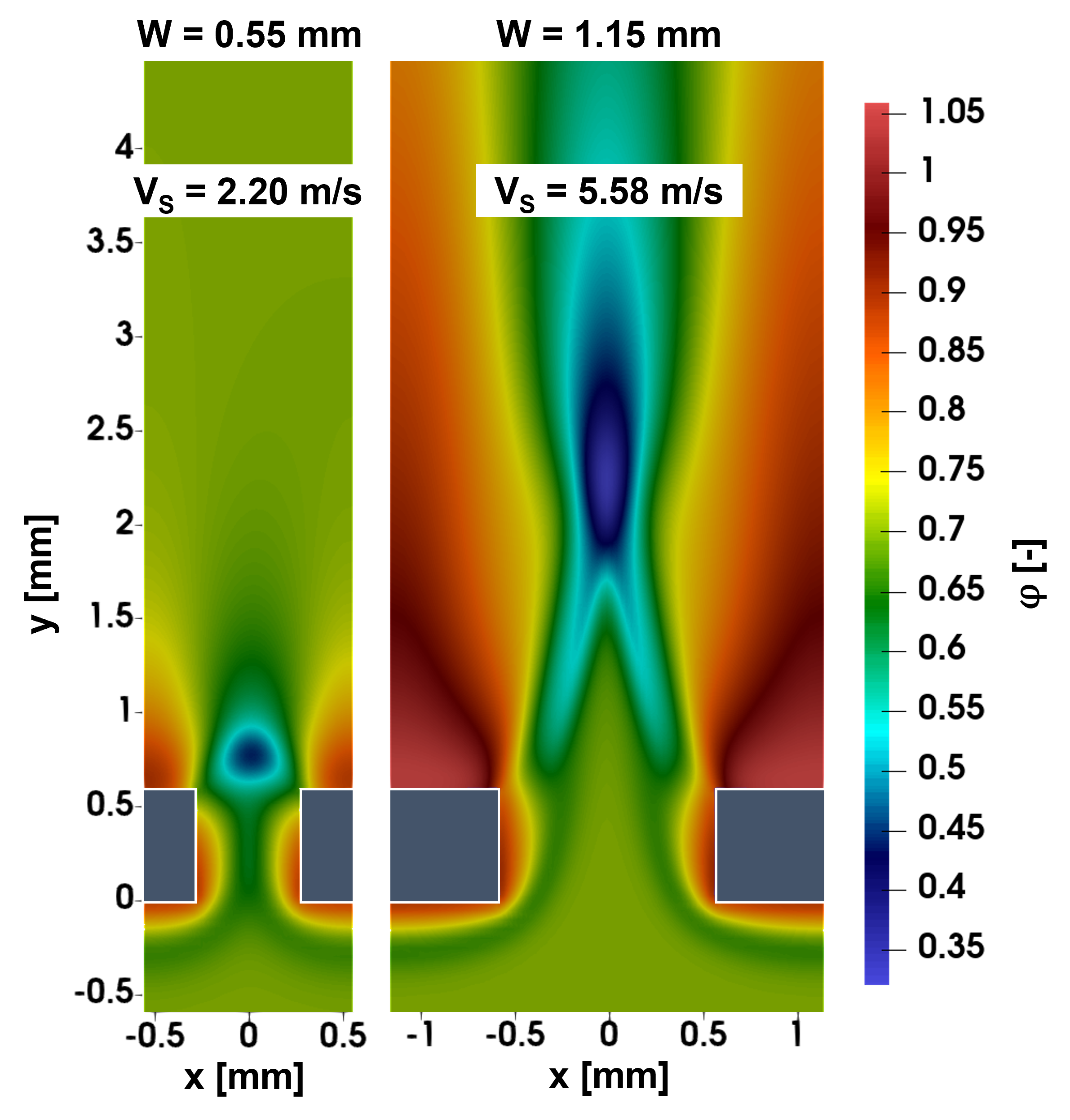}}
    \subfigure[]{\includegraphics[width=0.48\textwidth]{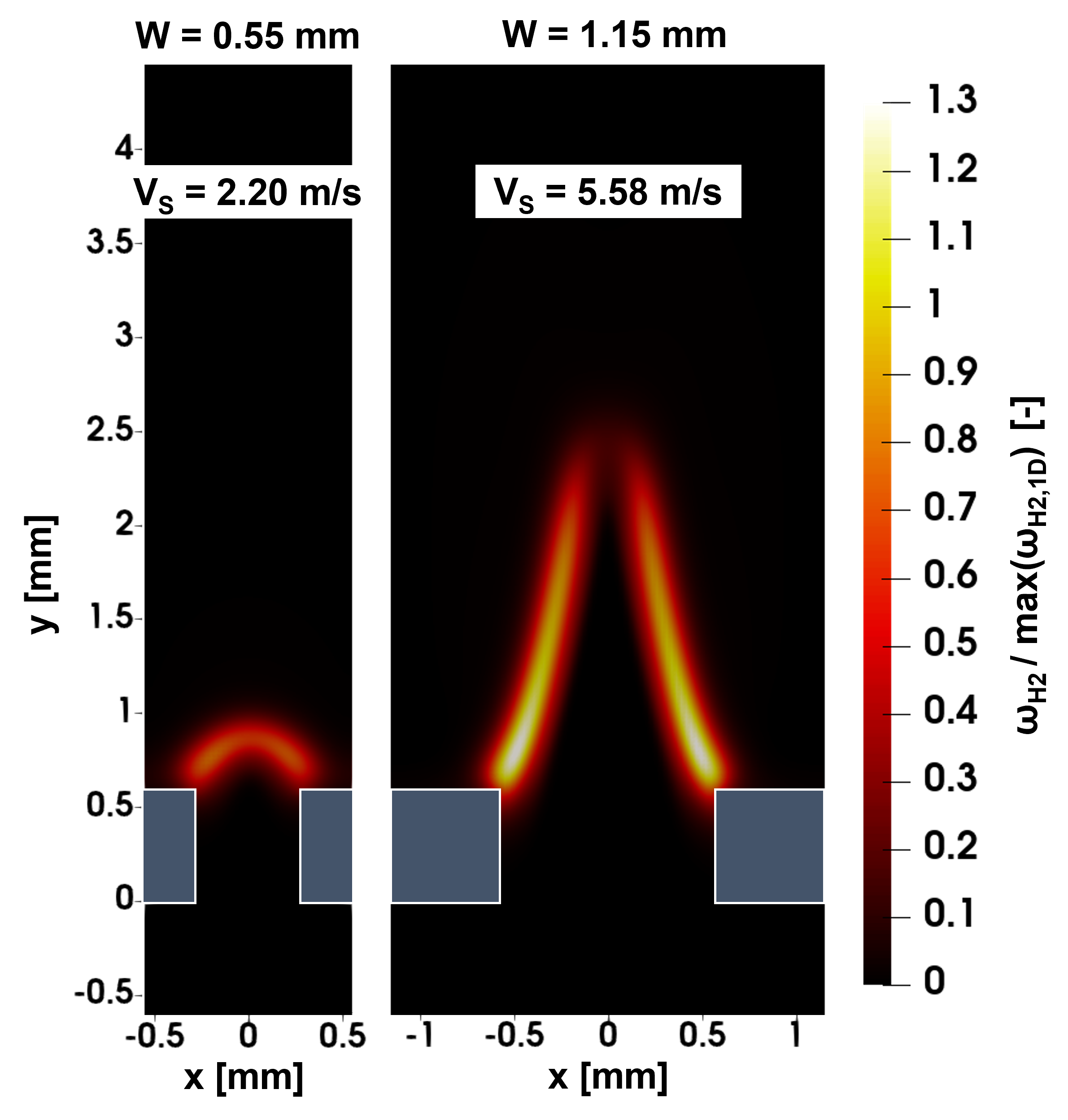}}
    \caption{(a) Local equivalence ratio for flashback limit cases for $W=\SI{0.55}{mm}$ and $W=\SI{1.15}{mm}$ with $\%\ch{H2}=98.9\%$ and $\phi=0.7$. (b) Normalized molecular hydrogen consumption rate for $W=\SI{0.55}{mm}$ and $W=\SI{1.15}{mm}$ with $\%\ch{H2}=98.9\%$ and $\phi=0.7$.}
    \label{fig:H2_W_contours}
\end{figure}

\begin{figure}[!ht]
\centering
\includegraphics[width=0.5\textwidth]{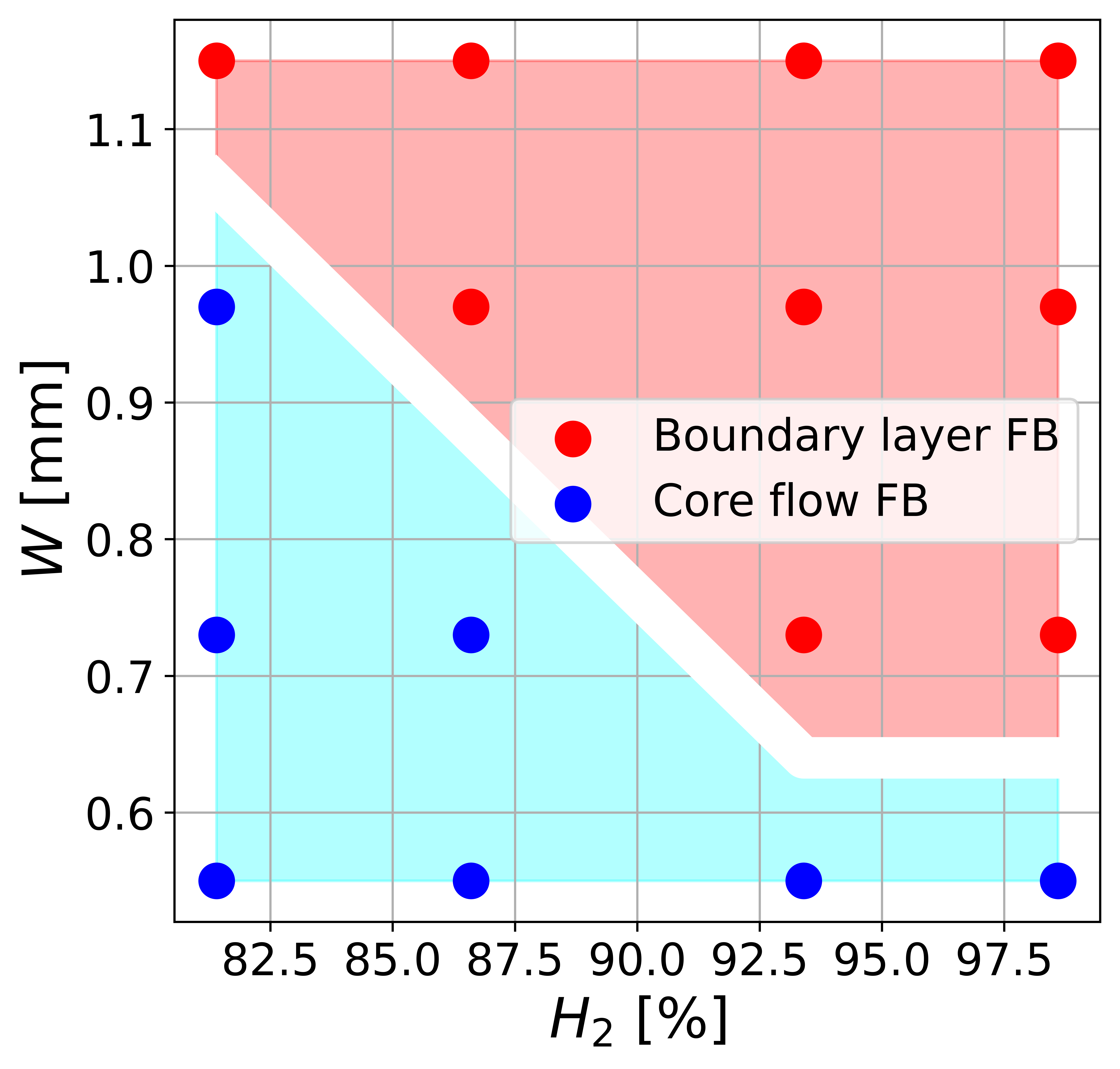}
\caption{Flashback regimes in the $\%\ch{H2}$ - $W$ parameter space. Blue represents ``core flow flashback", and red represents ``boundary layer flashback".}
\label{fig:border_H2_W}
\end{figure}

These results indicate that the slit width plays a crucial role in the physics of flame stabilization. At the same time, the substantial variability observed in $V_{FB}/s_L$ for high \ch{H2} contents suggests a significant influence of preferential diffusion effects. Fig.~\ref{fig:H2_W_contours} shows the local equivalence ratio, $\varphi$, and the normalized molecular \ch{H2} consumption rate, $\omega_{\ch{H2}}/\textnormal{max}(\omega_{\ch{H2},\textnormal{1D}})$, for the flashback limits of two cases with different slit widths, $W=\SI{0.55}{mm}$ and $W=\SI{1.15}{mm}$, and the same mixture, $\%\ch{H2}=98.9\%$ and $\phi=0.7$.
\noindent In the case of narrow slits, where the width is comparable to the flame thickness, the formation of highly curved flame front structures is limited, resulting in relatively weak preferential diffusion effects. Furthermore, the presence of low temperature gradients due to a uniform flow temperature at the slit exit suppresses Soret-induced preferential diffusion effects. On the other hand, wider slits with higher inlet velocities facilitate the development of highly curved flame fronts, leading to stronger preferential diffusion effects. Additionally, wider slits exhibit higher temperature gradients at the slit exit, promoting greater Soret-induced preferential diffusion effects in the flame base region. These combined effects contribute to a higher flashback velocity by enhancing the flame speed in the critical region close to the walls. In Fig.~\ref{fig:border_H2_W}, we show the parameter space divided into two regions corresponding to the two flashback regimes described above. It is evident that both parameters affect the flashback dynamics. We observe ``core flow flashback" in regions characterized by low values of $\%\ch{H2}$ and $W$, while ``boundary layer flashback" occurs in regions with high $\%\ch{H2}$ and $W$. Furthermore, upon comparison with the plot in Fig.~\ref{fig:H2_W_Vscaled}, we observe that the boundary between these two regimes coincides with $V_{FB}/s_L$ being between 1 and 2, suggesting that this may be a general trend within our modeling framework.

These results provide further confirmation of the significant role played by preferential diffusion in increasing the propensity for flashback when the hydrogen content is elevated. Moreover, we observe the crucial role of the slit width in influencing flame stabilization for high \ch{H2} content.

\paragraph{Burner temperature}

\begin{figure}[!ht]
    \centering
    \subfigure[]{\includegraphics[width=0.55\textwidth]{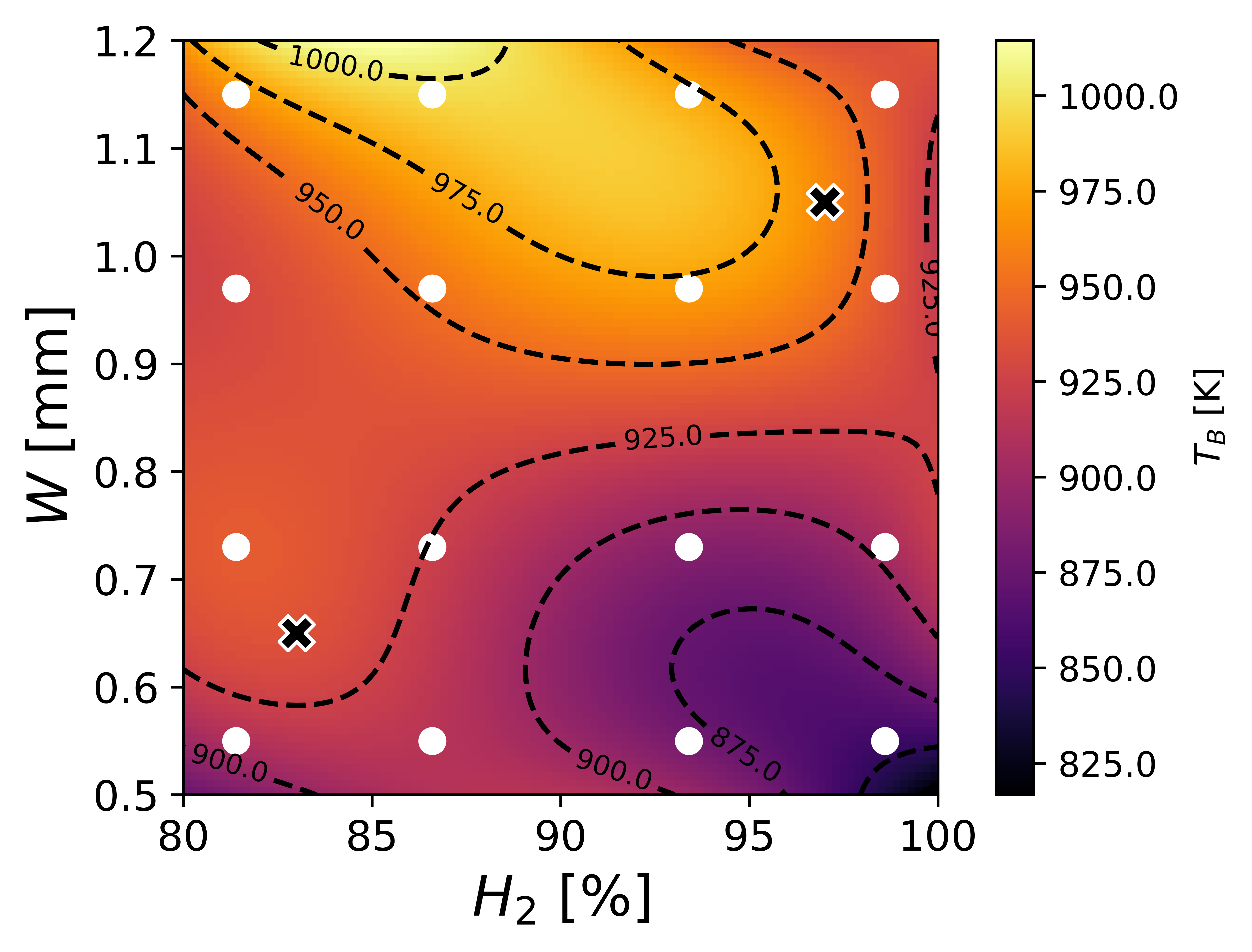}}
    \subfigure[]{\includegraphics[width=0.35\textwidth]{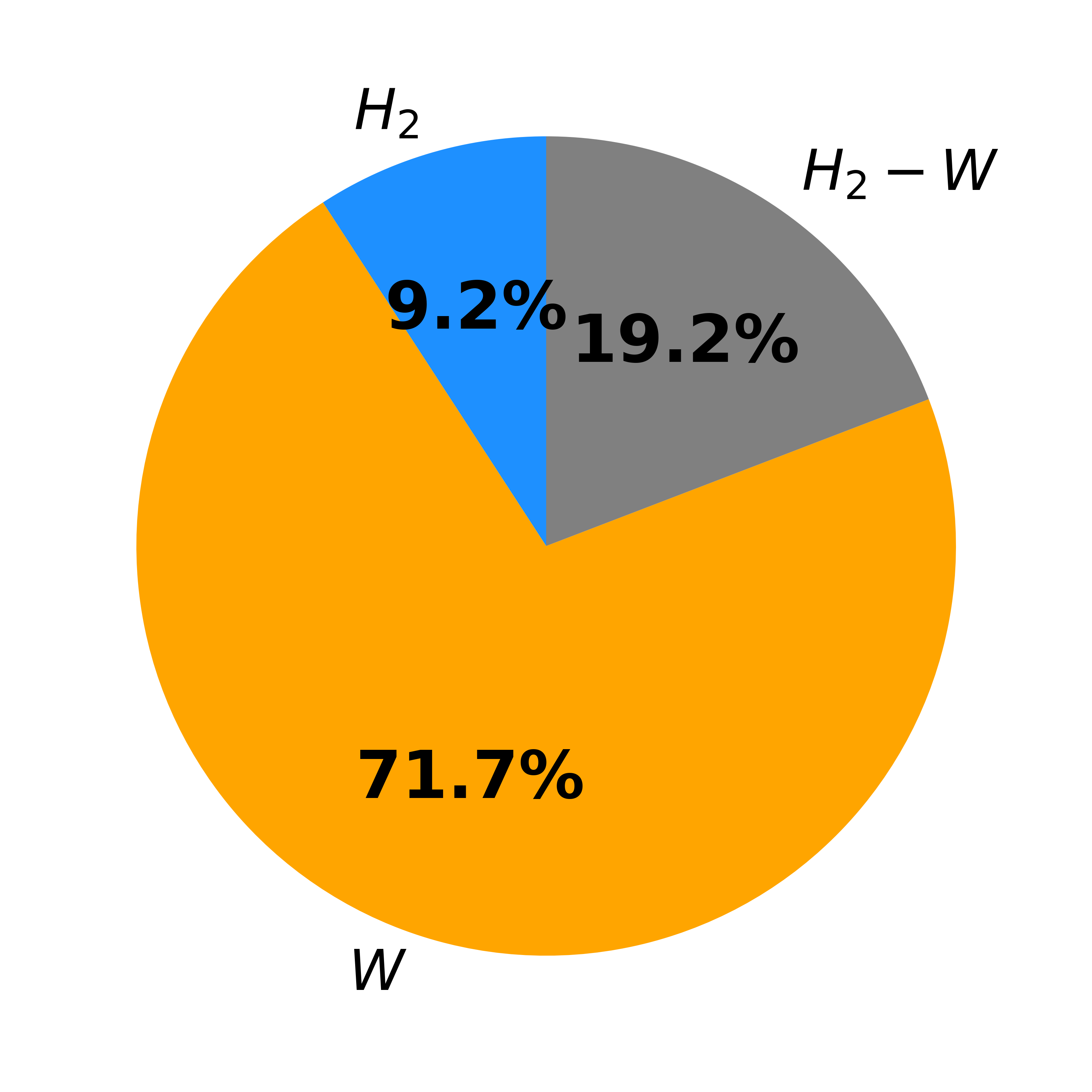}}
    \caption{(a) Stochastic response surface of $T_B$ in the $\%\ch{H2}$ - $W$ parameter space. White dots represent quadrature points. Black crosses indicate test points.} (b) Sobol' indices.
    \label{fig:H2_W_T}
\end{figure}

Fig.~\ref{fig:H2_W_T} shows the burner plate temperature map in the parameter space and the relative Sobol' indices.
\noindent We note a strong dependency of $T_B$ on the size of the slit, with wider slits resulting in higher temperatures. This is supported by the Sobol' index analysis, with $I_W$ indicating a dependence of 71.7\% on the slit width. Moreover, the interaction Sobol' index $I_{\ch{H2},W}$ reveals that the variation of $T_B$ within the investigated parameter range is not uniform: the maximum burner temperature, reaching approximately 1000 K, is observed in the region where ${W=\SI{1.2}{\mm}}$ and ${\%\ch{H2} \in [85\%,90\%]}$. Conversely, the minimum burner temperature of around 860 K is found in the region where ${\%\ch{H2} = 100\%}$ and ${W=\SI{0.5}{\mm}}$. 

It is important to note that in this study, increasing the slit width $W$ corresponds to an increase in the distance $D$ between two adjacent slits, to maintain a constant burner porosity. The work by Vance et al.~\cite{VANCEcorrelation} has demonstrated that changing the distance between the slits significantly impacts $T_B$ by influencing the heat transfer mechanisms between the burner, the flame, and the burnt gases. As a result, the variations in geometry, combined with different mixture characteristics, lead to complex interactions that significantly impact the heat transfer processes and, consequently, non-trivially influence the temperature of the burner.

\subsection{Effect of equivalence ratio and slit width}\label{sub:phiw}

Finally, the stochastic sensitivity analysis is performed on the last pair of parameters, i.e., the equivalence ratio and the slit width, in the ranges ${\phi\in[0.5,1.0]}$ and ${W\in[\SI{0.5}{\mm},\SI{1.2}{\mm}]}$, while fixing $\%\ch{H2}=100\%$. This specific value is chosen to maximize the sensibility to the parameters and due to its high practical significance. The test simulations are performed at the points corresponding to $\phi = 0.6$ and $W = \SI{1.05}{\mm}$, and $\phi = 0.9$ and $W = \SI{0.65}{\mm}$. Once again, the response surface values at these points align closely with the results from the test simulations, with an error margin of less than 2\%.

\paragraph{Flashback velocity}

\begin{figure}[!ht]
    \centering
    \subfigure[]{\includegraphics[width=0.55\textwidth]{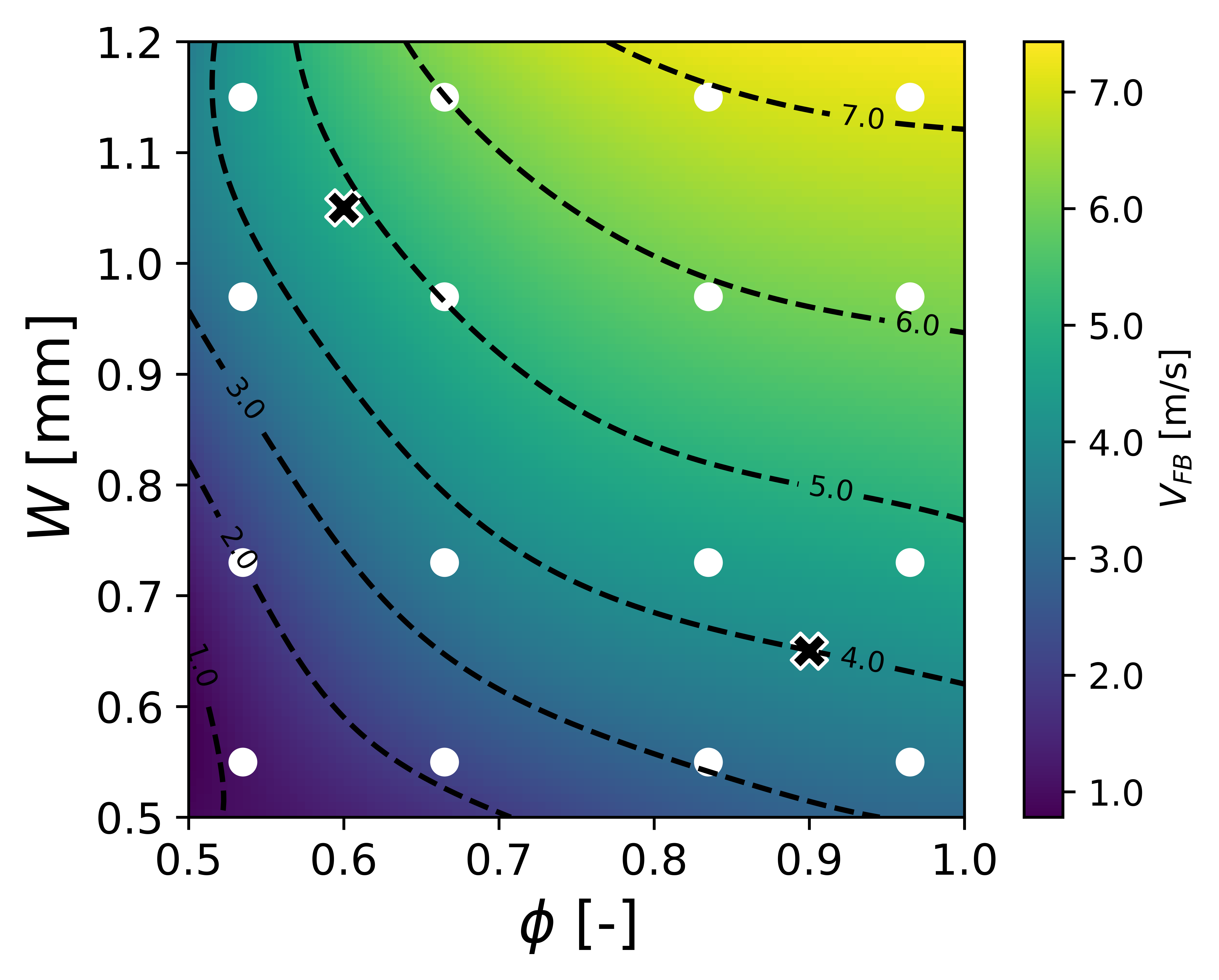}}
    \subfigure[]{\includegraphics[width=0.35\textwidth]{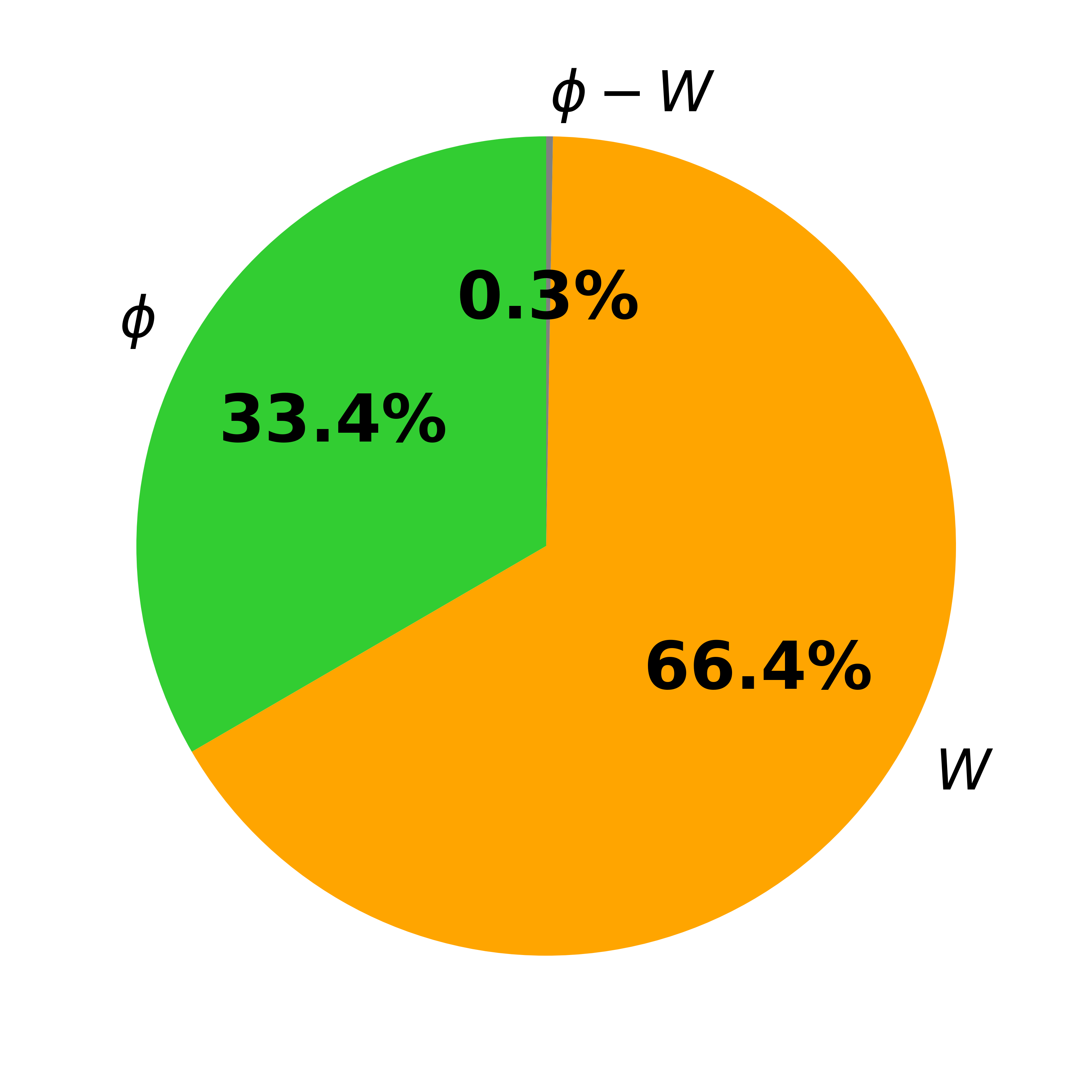}}
    \caption{(a) Stochastic response surface of $V_{FB}$ in the $\phi$ - $W$ parameter space. White dots represent quadrature points. Black crosses indicate test points.} (b) Sobol' indices.
    \label{fig:phi_W_V}
\end{figure}

In Fig.~\ref{fig:phi_W_V}, we show the flashback velocity as a function of the two input parameters and the corresponding Sobol' indices.
Even when keeping the \ch{H2} content constant, a significant variability in $V_{FB}$ is observed, ranging from approximately 1.0 m/s to 7.0 m/s. The primary parameter influencing this variability is $W$, as indicated by the Sobol' index $I_W=66.4\%$. Additionally, $\phi$ contributes to the variation of $V_{FB}$, with a Sobol' index of 33.4\%. Notably, the influence of these parameters on the flashback velocity is independent of each other, as indicated by the null interaction Sobol' index.

\paragraph{Normalized flashback velocity}

Fig.~\ref{fig:phi_W_V} shows $V_{FB}/s_L$ in the input parameters space and the relative Sobol' indices.
\begin{figure}[!ht]
    \centering
    \subfigure[]{\includegraphics[width=0.55\textwidth]{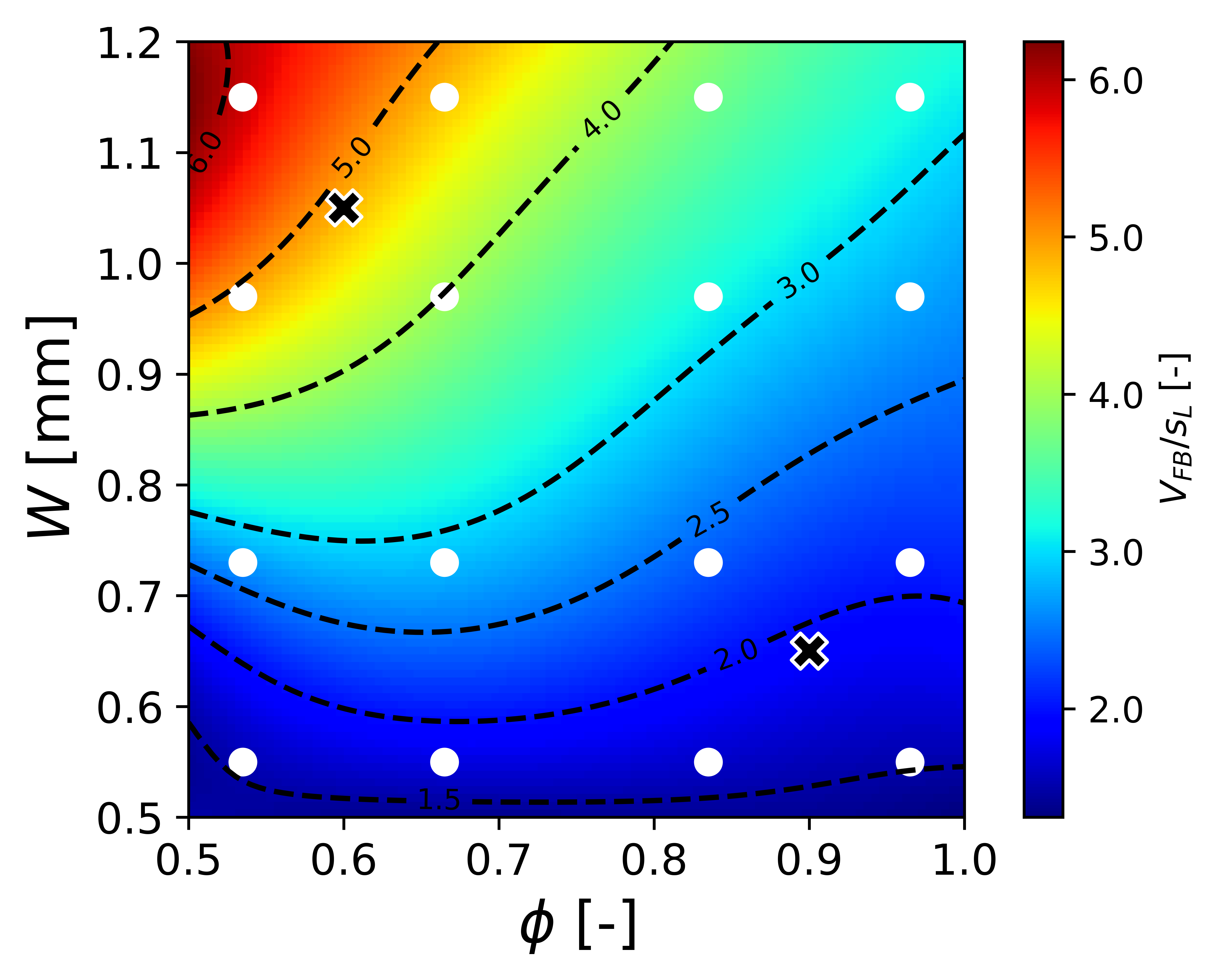}}
    \subfigure[]{\includegraphics[width=0.35\textwidth]{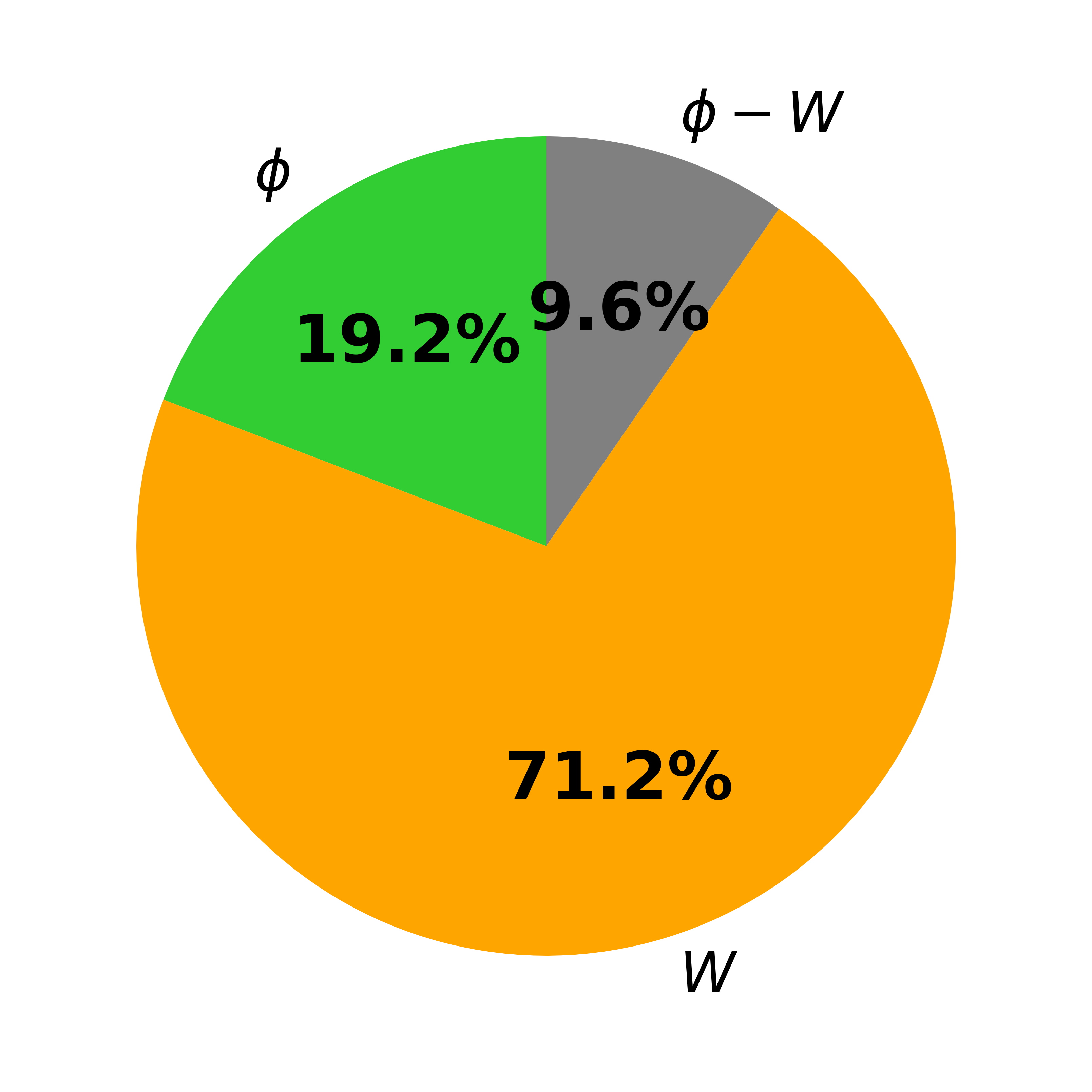}}
    \caption{(a) Stochastic response surface of $V_{FB}/s_L$ in the $\phi$ - $W$ parameter space. White dots represent quadrature points. Black crosses indicate test points.} (b) Sobol' indices.
    \label{fig:phi_W_Vscaled}
\end{figure}
\noindent Once again, significant variability is observed in the output quantity within the parameter space, with $V_{FB}/s_L$ ranging from 1.2 to 6.1. The maximum value is observed for larger $W$ and lower $\phi$. The normalized flashback velocity is influenced by 71.2\% by the slit width and by 19.2\% by the equivalence ratio. Furthermore, the interaction Sobol' index indicates that the variability of $V_{FB}/s_L$ is not uniform, with a higher impact when varying the slit width for low equivalence ratios.

\begin{figure}[!ht]
\centering
\includegraphics[width=0.5\textwidth]{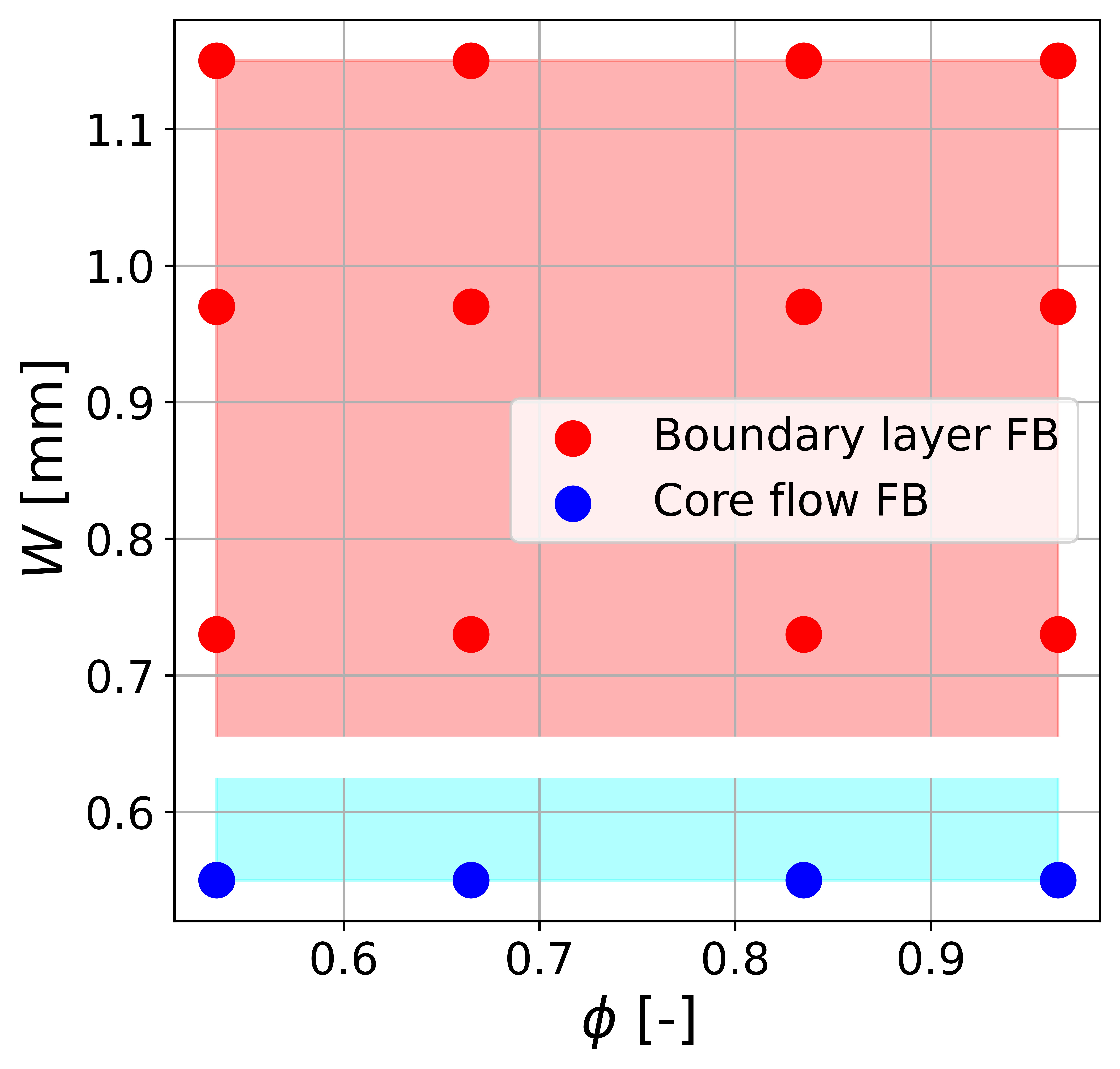}
\caption{Flashback regimes in the $\phi$ - $W$ parameter space. Blue represents ``core flow flashback", and red represents ``boundary layer flashback".}
\label{fig:border_phi_W}
\end{figure}

As expected from the previous analysis, which highlighted the influence of preferential diffusion effects on the relationship between $V_{FB}/s_L$ and $W$, we observe a pronounced effect of the slit width, particularly for low equivalence ratios where preferential diffusion effects are more prominent. Another perspective is that the ratio $V_{FB}/s_L$ exhibits a significant dependence on the equivalence ratio when the slit is sufficiently wide to allow preferential diffusion effects to play a crucial role. It is important to note that the extent of this influence may also vary depending on the chosen hydrogen content: in this case, being $\%\ch{H2}=100\%$, we assume that we are observing the maximum possible variability attributed to preferential diffusion effects. The division of the parameter space into ``core flow flashback" and ``boundary layer flashback" regimes is presented in Fig.~\ref{fig:border_phi_W}. Again, it can be observed that the transition corresponds to values of $V_{FB}$ between 1 and 2, reaffirming the universality of this phenomenon in our simulation context.

\paragraph{Burner temperature}

\begin{figure}[!ht]
    \centering
    \subfigure[]{\includegraphics[width=0.55\textwidth]{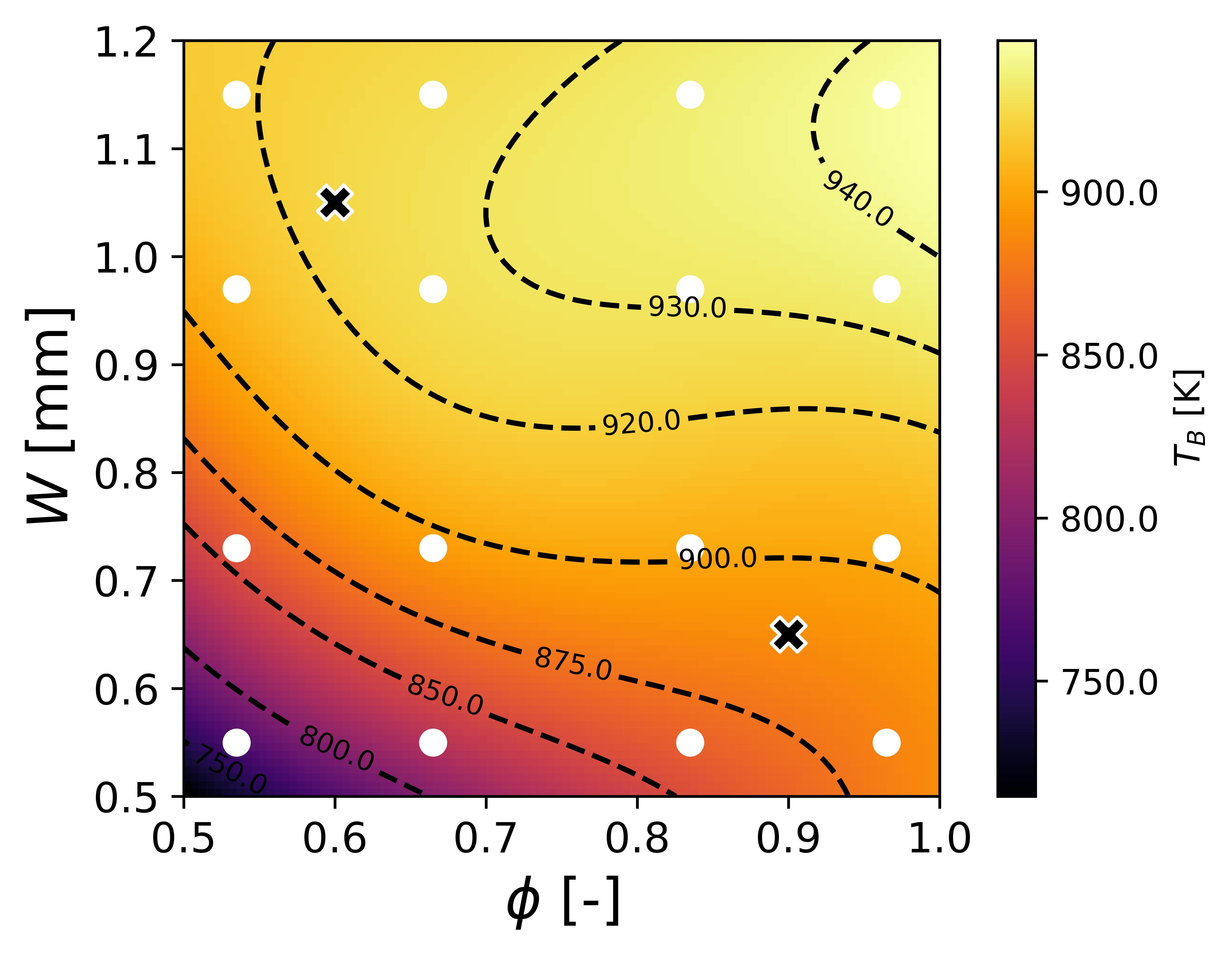}}
    \subfigure[]{\includegraphics[width=0.35\textwidth]{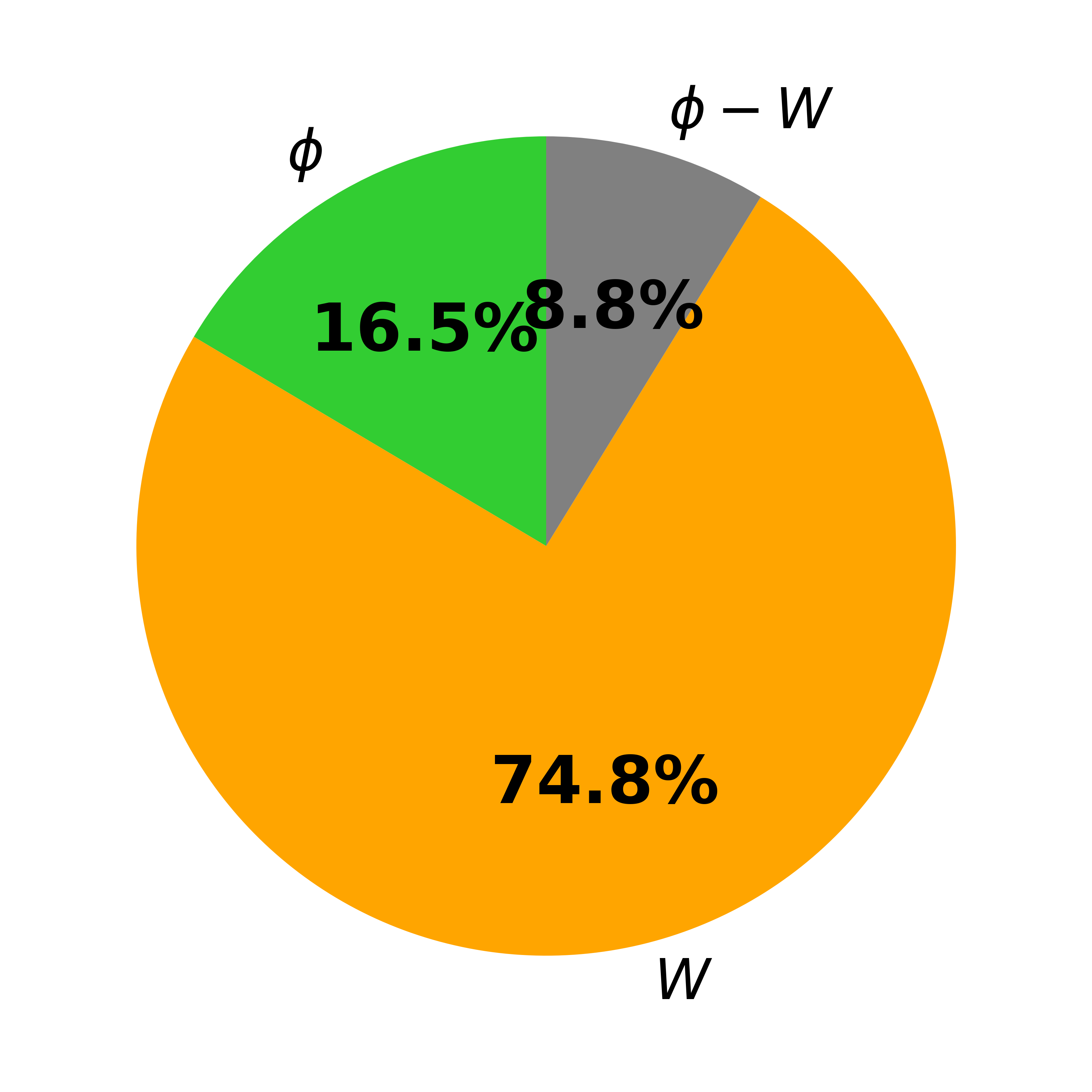}}
    \caption{(a) Stochastic response surface of $T_B$ in the $\phi$ - $W$ parameter space. White dots represent quadrature points. Black crosses indicate test points.} (b) Sobol' indices.
    \label{fig:phi_W_T}
\end{figure}

The variation of the burner plate temperature $T_B$ in the domain of interest is shown in Fig.~\ref{fig:phi_W_T} with the relative Sobol' indices.
\noindent We observe the burner temperature varying uniformly over the parameter space, with $T_B$ increasing when increasing both $\phi$ and $W$. The Sobol' indices indicate a stronger dependence on $W$, being $I_W=74.8\%$ and $I_\phi=16.5\%$. We note that the burner plate temperature varies considerably on the domain, with an increase of $\sim 20\%$ going from the lower-left corner ($T_B\simeq\SI{760}{K}$) to the upper-right corner ($T_B\simeq\SI{940}{K}$) in Fig.~\ref{fig:phi_W_T}(a).

\section{Conclusions}

In this work, the use of gPC and stochastic sensitivity analysis has allowed us to gain a comprehensive and deep insight into the geometrical and operating parameters affecting the occurrence of flashback in perforated burners. The flashback velocity, $V_{FB}$, is primarily influenced by the \ch{H2} content in the mixture. As the \ch{H2} content increases, the flashback velocity also increases. However, as the mixture approaches 100\% \ch{H2}, the effects of the equivalence ratio and the slit width become more significant, with $V_{FB}$ decreasing with $\phi$ and increasing with $W$. It is important to note that the variability of $V_{FB}$ is closely linked to the variability of the laminar flame speed $s_L$ of the specific mixture under consideration. Therefore, $V_{FB}$ may not provide a valid measure of the intrinsic flashback propensity of the mixture. To eliminate the dependence on the laminar flame speed, the normalized value of the flashback velocity, $V_{FB}/s_L$, is considered as a reliable indicator to evaluate the flashback propensity of a given parameter combination. Interestingly, regions in the parameter space that are more susceptible to preferential diffusion effects, such as high \ch{H2} content and low equivalence ratio, exhibit higher values of $V_{FB}/s_L$. This is because preferential diffusion effects, particularly Soret-induced and curvature-induced preferential diffusion, play a significant role in flame stabilization. Furthermore, the slit width also plays a crucial role in determining the flashback propensity, especially when the hydrogen content is high or the equivalence ratio is low. All of the parameters under consideration exert an influence on the flashback dynamics, leading a transition between the ``core flow flashback" and ``boundary layer flashback" regimes as they are varied. Notably, this transition consistently occurs in the range of $V_{FB}/s_L\lesssim1.5$ to $V_{FB}/s_L\gtrsim2$. The burner temperature at the flashback limit, $T_B$, is mainly influenced by the burner geometry. Specifically, wider slits result in higher burner temperatures. The variability of $T_B$ in the parameter space can be attributed to several factors, including the position of the flame at the flashback limit, the total mass flow rate, radiative losses, the attachment of the flame to the burner plate, and the contact area between the hot burnt gases and the top surface of the burner plate. The aforementioned factors are dependent on both the specific geometry of the burner and the characteristics of the mixture, such as the adiabatic flame temperature, flame thickness, and Lewis number. 

It is important to acknowledge that in practical devices, the burner temperatures can be affected by factors such as the finite size of the burners, necessitating potential corrections and adaptations of the model for real-world implementation. Furthermore, while our current model is based on a 2D representation, it is essential to consider that the three-dimensional structure of the slits could introduce significant variations in flashback dynamics, which may warrant further investigations. Nevertheless, our study sheds light on the complex interplay of various parameters and their interactions that influence the flashback propensity in high hydrogen content mixtures, with particular emphasis on the crucial role of preferential diffusion effects. In future works, further analyses could delve into understanding the relative impact of different sources of preferential diffusion on flashback velocity, including the Soret effect. The insights gained hold the potential for guiding the design and optimization of perforated burners, ensuring their safe and efficient operation in real-world applications. It should be noted that to achieve this, calibration and validation data from experiments, which are currently lacking, would be necessary.

\section*{Author contributions}

Filippo Fruzza and Rachele Lamioni: Conceptualization, Methodology, Data curation, Writing-Original draft preparation. Maria Vittoria Salvetti and Alessandro Mariotti: Methodology, Reviewing and Editing. Chiara Galletti: Conceptualization, Methodology, Writing-Original draft preparation.  

\section*{Acknowledgements}

This research is funded by the Ministry of University and Research (MUR) and Immergas S.p.A., Brescello, RE (Italy),  as part of the PON 2014-2020  ``Research and Innovation" resources - Green/Innovation Action - DM MUR 1061/2021 and DM MUR 1062/2021. We also thank Ing. Cristiana Bronzoni and Ing. Marco Folli from Immergas S.p.A. for the valuable discussions.

\bibliography{fuel_flashback}

\end{document}